\documentclass[aps,twocolumn,prd,superscriptaddress]{revtex4-2}

\usepackage[utf8]{inputenc}

\usepackage{mathtools}
\usepackage{amsfonts}
\usepackage{mathrsfs}
\usepackage{bm}
\usepackage{bbm}
\usepackage{slashed}
\usepackage{tensor}
\usepackage{listings}
\usepackage{graphicx}
\usepackage{color}
\usepackage[dvipsnames]{xcolor}
\usepackage{array}
\usepackage[abs]{overpic}

\usepackage{placeins}

\usepackage{makecell}
\usepackage{subcaption}

\usepackage{xspace}
\usepackage{siunitx}
\usepackage{xfrac}
\usepackage{hyperref}
\usepackage[nameinlink]{cleveref}
\usepackage{appendix}

\usepackage{xifthen}
\usepackage{xcolor}
\hypersetup{
	colorlinks,
	linkcolor={red!75!black},
	citecolor={blue!75!black},
	urlcolor={blue!75!black}
}

\usepackage{booktabs}
\usepackage{multirow}

\newcolumntype{C}{>{$}c<{$}}
\AtBeginDocument{
	\heavyrulewidth=.08em
	\lightrulewidth=.05em
	\cmidrulewidth=.03em
	\belowrulesep=.65ex
	\belowbottomsep=0pt
	\aboverulesep=.4ex
	\abovetopsep=0pt
	\cmidrulesep=\doublerulesep
	\cmidrulekern=.5em
	\defaultaddspace=.5em
}

\newcommand{\imag}{\text{i}}
\newcommand{\tinytext}[1]{\text{\tiny{#1}}}

\newcommand{\Tr}{\ensuremath{\operatorname{Tr}}}

\newcolumntype{L}{>{\centering\arraybackslash}m{3cm}}

\graphicspath{{./figures/}}

\newcommand{\gettitle}{Flowing fields and optimal RG-flows}

\newcommand{\getHeidelbergAffiliation}{\affiliation{Institut f{\"u}r Theoretische Physik, Universit{\"a}t Heidelberg, Philosophenweg 16, 69120 Heidelberg, Germany}}
\newcommand{\getEMMIAffiliation}{\affiliation{ExtreMe Matter Institute EMMI, GSI, Planckstr. 1, 64291 Darmstadt, Germany}}

\hypersetup{
	pdftitle={\gettitle},
	pdfauthor={Ihssen},
	pdfkeywords=
	{functional renormalization group} {effective potential},
	bookmarksopen=true,
	bookmarksopenlevel=2,
	bookmarksnumbered=true
}

\begin{document}

\title{\gettitle}

\author{Friederike Ihssen}\getHeidelbergAffiliation
\author{Jan M. Pawlowski}\getHeidelbergAffiliation\getEMMIAffiliation

\begin{abstract}
Renormalisation group approaches are tailor made for resolving the scale-dependence of quantum and statistical systems, and hence their phase structure and critical physics. Usually this advantage comes at the price of having to truncate the full theory at hand, which asks for optimal expansion schemes. In the present work we use a functional renormalisation group (fRG) approach for the effective action which includes general scale-dependent reparametrisations of the theory \cite{Pawlowski:2005xe}. This approach is used in an O(N)-theory to set up adaptive RG-flows that correspond to an optimal systematic expansion of the theory about the ground state or rather its full covariance or propagator. These parametrisations are induced by flowing fields that encode the differential reparametrisation steps. The approach is put to work for an investigation of the thermal phase transition in the O(4)-theory in view of applications to QCD. The respective results are compared with those obtained in standard fRG computations. 
\end{abstract}

\maketitle

\section{Introduction}
\label{sec:Introduction}

Renormalisation group approaches are ideally suited for resolving the intricate phase structure of physical systems, ranging from condensed matter and statistical systems over Quantum Chromodynamics, particle physics, cosmology to quantum gravity. They are set-up for monitoring the change of physical phenomena and degrees of freedom under changing intrinsic scales such as mass and more generally couplings, as well as external parameters such as temperature, density or external fields. Renormalisation group approaches provide a relatively easy access to the mechanisms or dynamics behind intricate physical phenomena already in simple approximations of the system at hand, even for strongly correlated systems. Often, simple approximations already lead to semi-quantitative results. 

In turn, fully quantitative results and, equally important, a small systematic error estimate require far more effort. Moreover, the phase structure of many systems, including many condensed matter systems and QCD, exhibits regimes with competing order effects. There, quantitative precision as well as the inclusion of the full vacuum structure is mandatory for obtaining even qualitative results. Accordingly, the quest for quantitative precision and small systematic error estimates is of vital importance for the reliability of the RG-approach, and the present work aims at adding to this important endeavour. 

A specifically appealing feature of RG-approaches is their adaptive flexibility that originates in the successive resolution of the theory at hand, and Wilsonian fRG approaches are tailor made in this respect. In the present work we aim at a description of the one particle irreducible (1PI) effective action of free energy $\Gamma$ in terms of the emergent dynamical degrees of freedom. This corresponds to an expansion of the theory about its ground state, and in terms of the effective action this is tantamount to an expansion about the full covariance or inverse propagator. In the fRG approach this idea is implemented within a successive RG-adapted reparametrisation of the dynamical field, which evolves from the microscopic field at large momentum scales to the emergent dynamical field at low momentum scale. This requires an fRG flow equation for the effective action that encompasses general reparametrisations, which has been set-up in \cite{Pawlowski:2005xe, Salmhofer:2006pn}, for a partial reparametrisation setup see \cite{Wetterich:1996kf}. The underlying idea dates back to \cite{Wegner_1974} and has recently also been used for the construction of the essential fRG, \cite{Baldazzi:2021ydj, Baldazzi:2021orb}.

In \Cref{sec:FlowingFields} we introduce the general flow equation that allows for general reparametrisations (flowing fields), and discuss systematic expansion schemes as well as systematic error estimates. In \Cref{sec:GenRG} we apply the flowing fields approach to O(N) models including a discussion of qualitatively different flowing transformations. In the current work we use the flowing field approach to reduce the full dispersion of the field to a classical one at each RG-step. Numerical results for the three-dimensional O(4)-theory in the broken phase as well as for the thermal phase transition of the O(4)-theory in four dimensions are presented in \Cref{sec:results}. We summarise our findings in \Cref{sec:summary}.

\section{Systematic expansions, systematic error estimates and flowing fields}
\label{sec:FlowingFields}

In this section we discuss systematic expansion schemes for the functional renormalisation group (fRG) approach and their optimisation via generic field reparametrisations during the flow. In \Cref{sec:GenFlowGen} we discuss the flow equation for the effective action in its standard form \labelcref{eq:Flow} and the general flow equation with flowing fields \labelcref{eq:GenFlow}. This is accompanied with a brief overview of common expansion schemes. 

In \Cref{sec:SysExpandQCD} we discuss expansion schemes and their systematics within QCD, which is one of the main application we have in mind for the present approach. Moreover, it is one of the few theories where most of the standard expansion schemes have been applied, in most cases in combination. These applications also come together with systematic error estimates. 

In \Cref{sec:FlowingPhi} we discuss the use of flowing fields in the general flow \labelcref{eq:GenG2} for an RG-adapted expansion scheme at two qualitatively different examples. In this context it is convenient to distinguish three different construction principles behind the choice of flowing fields:  
\begin{itemize} 
\item[(i)] Firstly, we can use flowing fields to adapt the field basis to the phase (symmetric or broken) the theory is in, at a given cutoff scale $k$. This application of the generalised flow  in \cite{Pawlowski:2005xe} has been put forward in \cite{Lamprecht2007} in an O(N)-theory. It has been called Goldstonisation, and is a scale-dependent non-linear transformation in field space. Naturally, this optimises a given expansion scheme. An application directly based on \cite{Lamprecht2007} can be found in \cite{Isaule:2018mxt, Isaule:2019pcm}, for a linear version see \cite{Daviet:2021whj}. 
\item[(ii)] Flowing fields can be used to enforce the expansion of the effective action about the full two-point function or covariance of the theory. This powerful idea has been put forward in \cite{Salmhofer:2006pn}, for a recent application to complex effective actions see \cite{Ihssen:2022xjv}. This choice is also enforced by full functional optimisation as 
put forward in \cite{Pawlowski:2005xe}. It is also obtained within the implementation of optimal transport \cite{Cotler:2022fze}, applied to the Wegner flow \cite{Wegner_1974}, or the flow of the divergence functional, \cite{Floerchinger:2023qpw}. 
\item[(iii)] On the other hand, we can use flowing fields for simplifying the effective action, which is at the root of the essential RG put forward in \cite{Baldazzi:2021ydj}. There, these simplifications are classified in terms of essential and inessential couplings \cite{Wegner_1974}. 
\end{itemize}
In all these cases, a reparametrisation with \textit{flowing} fields allows us to expand the theory, or rather its effective action, about a dynamically adjusted expansion point that is as close as possible to the scale-dependent ground state of the theory. Note also, that a generic use of flowing fields, that aims at optimal expansions, is almost inevitably a mixture of \textit{(i)-(iii)}. 

In \Cref{sec:FlowingPhi} we specifically discuss examples for \textit{(i)} and \textit{(iii)}. Moreover, the explicit application in the present work falls into the class \textit{(iii)}, but implicitly also enforces \textit{(ii)}. We put the flowing fields to work in the fRG approach of the one-particle irreducible (1PI) effective action. The general flow in the presence of flowing fields used in this work has been derived in \cite{Pawlowski:2005xe} and is discussed later in \Cref{sec:GenRG}.

\subsection{General flows for the effective action}
\label{sec:GenFlowGen}

For the discussion in the present section we start with the standard 1PI flow, the Wetterich equation \cite{Wetterich:1992yh}. For a scalar theory with a real scalar field $\varphi$, the flow for the effective action $\Gamma_k[\varphi]$ reads 
\begin{align}
	\partial_t \Gamma_k[\varphi] =\frac{1}{2} \int\frac{d^d p}{(2\pi)^d}\frac{1}{\Gamma^{(2)}_k[\varphi] +R_k}(p,-p)\,\partial_t R_k(p) \,, 
	\label{eq:Flow} 
\end{align}
with an infrared cutoff scale $k$. The infrared cutoff is introduced via the regulator function $R_k$, which is added to the classical dispersion. The choice used here is detailed in \Cref{app:regulator}. Roughly speaking it introduces an IR mass which vanishes for $k \to 0$. Hence, quantum fluctuations below the cutoff scale $p^2 \lesssim k^2$ are suppressed and are integrated out by lowering $k$. This is implemented via the logarithmic scale derivative w.r.t.~$t= -\log (k/k_\textrm{ref})$, with some reference scale $k_\textrm{ref}$, in contrast to the standard definition without the minus sign. The minus sign leads to the natural interpretation of $t$ as an RG-'time': in the numerical solution scheme used here, the RG-flows are formulated as convection-diffusion equations. In the present work we choose the reference scale to be the UV-cutoff scale of the flow, $k_\textrm{ref}=\Lambda$, and hence the RG-time $t$ runs from 0 to $\infty$. The physical theory is approached for $k\to 0$ and the physical effective action is given by $\Gamma[\varphi]=\Gamma_{k=0}[\varphi]$. 

\Cref{eq:Flow} has a simple one-loop form in terms of the full field- and momentum-dependent propagator, and its one loop form readily extends to general theories with a general field content. Then, $\varphi$ is a superfield, that comprises all fundamental fields of the theory at hand, and $R_k$ is a regulator matrix in field space. Moreover, the momentum integral turns into a trace over internal momenta, Lorentz indices and species of fields, including minus signs for fermions. 

For example, in QCD we have $\varphi=(A_\mu, c, \bar c, q, \bar q)$, where the gluons fields are denoted $A_\mu$, the ghosts $c$ and quarks $q$. The flows of its $n$-th moments, the one-particle irreducible correlation functions, are derived from this master equation by taking $n$ field derivatives of the flow of $\Gamma[\varphi]$, 
\begin{align}
\Gamma_{\varphi^{i_1}\cdots \varphi^{i_n}}^{(n)}[\varphi](p_1,...,p_n)=	\frac{\delta^n\Gamma[\varphi]}{\delta\varphi_{i_1}(p_1)\cdots\delta\varphi_{i_n}(p_n) }\,, 
\label{eq:GnNotation}
\end{align}
or in short $\Gamma_{\varphi^{i_1}\cdots \varphi^{i_n}}$ or $\Gamma_{i_1\cdots i_n}$ for increasing the readability of expressions. This leads us to a coupled infinite hierarchy of flows for these moments or correlation functions, which typically only admits analytical or numerical solutions within a given order of a systematic expansion scheme of the full system at hand. 

One of these systematic expansion schemes is the vertex expansion, where the effective action is expanded in terms of its moments, while keeping the full momentum dependences of these moments. A complimentary systematic scheme is the derivative expansion, where all moments are taken into account, but their momentum dependence is expanded in powers of momenta. Evidently, these schemes can be combined in a systematic way, leading to mixed schemes.   

All expansion schemes have in common that the respective expansion of the effective action is linked to an expansion of the theory, or its Hilbert space, around a state that captures the essential features of the full ground state. This is most evident within the derivative expansion where the zeroth order takes into account point like interactions to all orders without any momentum dependences in form of an effective potential. Then, higher orders include momentum dependences in powers of $p^2/m_\textrm{gap}^2$, where $m_\textrm{gap}^2$ is a characteristic mass scale of the theory. Evidently, such a scheme fails (or shows a rather slow convergence) in the presence of intricate momentum or angular dependences of scattering processes. One such example are complicated Fermi surfaces in interacting theories. 

In turn, the vertex expansion is an expansion in the full moments of the theory, starting with the momentum-dependent (inverse) propagator, the full covariance of the theory. Its implicit expansion parameter is the phase space suppression of higher order interactions for local interactions, and the one loop form of the Wetterich flow gives an easy access to the respective error control. While it is amiable towards intricate momentum dependences of vertices, the convergence of this scheme is slowed down in the presence of long range resonant interactions or large couplings. This can be treated in terms of emergent dynamical effective fields that are introduced for resonant interaction channels within the flow equation. For four-field interactions such as four-quark scatterings in QCD this has been developed in \cite{Gies:2001nw}. It is tantamount to a flowing self-consistent Hubbard-Stratonovich transformation and are simple to implement due to the underlying Gau\ss ian nature of the field transformation. 

The fRG approach for 1PI flows with general (non-Gau\ss ian) field reparametrisation or flowing fields has been developed in \cite{Pawlowski:2005xe}. It incorporates the a general reparametrisation $\varphi\to \phi_k[\varphi]$. Importantly, its use only requires the explicit choice of the differential change of the composite field $\dot \phi_k[\phi]$ at a given cutoff scale $k$ in terms of the composite field basis $\phi_k$ at this scale. The respective flow of the effective action of the dynamical composite or flowing field $\phi$ is given by  
\begin{align}
	\left( \partial_t + \int_x \dot{\phi} \frac{\delta}{\delta \phi} \right) \Gamma_k[\phi] =\frac{1}{2} \Tr\left[ G[\phi]\left(\partial_t + 2 \frac{\delta \dot{\phi}}{\delta \phi} \right) \, R_k\right] \,, 
	\label{eq:GenFlow} 
\end{align}
where $G[\phi]$ is the full, field- and momentum-dependent propagator of the theory, 
\begin{align}
	G_{\phi_i \phi_j}[\phi](p,q) = \left[ \frac{1}{\Gamma_k^{(2)}+R_k}\right]_{\phi_i \phi_j}(p,q)\,,
	\label{eq:Prop}
\end{align}
also using the short-hand notation $G_{i j}$. The regulator matrix $R_k$ is added to the classical dispersion, again the choice used here is detailed in \Cref{app:regulator}. 

We emphasise again, that the differential flow $\dot \phi$ in \labelcref{eq:GenFlow} is at our disposal. For example, we may choose and explicit field transformation in the class \textit{(i)}, as put forward in \cite{Lamprecht2007, Isaule:2018mxt, Isaule:2019pcm, Daviet:2021whj} and discussed in \Cref{sec:FlowingPhi} as one of the examples. Alternatively we may enforce the class \textit{(ii)} or \textit{(iii)} with constraints on the flow, e.g.~that of the two-point function or a specific momentum channel of a four-point function. Structurally, this is given by  
\begin{align}
	\dot\phi_k(\phi,k) \qquad \longrightarrow \qquad \partial_t \Gamma_k^{(n)}[\phi](p)\equiv 0\,, 
	\label{eq:FlowingFieldsGen} 
\end{align}
for a specific $n$ and $\phi\in {\cal I}_\phi$, $p\in {\cal I}_p$ take values in some intervals or sets ${\cal I}_\phi$ and $ {\cal I}_p$ respectively. \Cref{eq:FlowingFieldsGen} entails the differential form of the transformation from the microscopic fundamental fields to emergent effective fields. So far, it has been considered for $n=2$ in the context of functional optimisation in \cite{Pawlowski:2005xe} and \cite{Fischer:2008uz} with $p>k$, in \cite{Salmhofer:2006pn} for keeping the covariance of the theory fixed, and in \cite{Baldazzi:2021orb, Baldazzi:2021ydj, Knorr:2022ilz, Knorr:2023usb} (for $p=0$) in the context of essential fRG flows. For $n=4$ is has been used in QCD for dynamical hadronisation of momentum channels of the four-quark scattering vertex, for further discussion see \Cref{sec:SysExpandQCD}. Moreover, it has been used for dynamical composite dimers in applications to ultracold gases in \cite{Floerchinger:2009pg}. In \Cref{sec:FlowingPhi} this general framework is illustrated within two examples, and the specific choice \labelcref{eq:FlowingFields} of \labelcref{eq:FlowingFieldsGen} is used for the explicit computations in the present work.

\subsection{Systematic expansions in QCD}
\label{sec:SysExpandQCD}

For an explanation of the tasks at hand we concentrate on QCD at finite temperature and density, which features the most advanced computations within mixtures of the two systematic expansion schemes discussed above: In QCD, mixtures of the vertex and derivative expansion with emergent dynamical fields for resonant channels of the four-quark interactions have been applied (dynamical hadronisation). In most cases only the scalar-pseudoscalar channel has been dynamically hadronised, leading to emergent pions and $\sigma$ mesonic fields, see \cite{Gies:2002hq, Braun:2009ewx, Floerchinger:2009uf, Mitter:2014wpa, Braun:2014ata, Cyrol:2017ewj, Fu:2019hdw}. For an additional dynamical hadronisation of the vector channel, leading to emergent vector mesons, see \cite{Rennecke:2015eba}. The complete structure of emergent hadrons has been discussed in \cite{Fukushima:2021ctq}. Augmented with a full effective potential for the respective mesonic fields, which includes the point-like scattering of the emergent mesonic fields for small momenta, this is a rather efficient description of QCD at high and low energies as well as at finite temperature and density or baryon-chemical potential. These applications can be considered as the first working examples for the power and efficiency of the flowing fields approach. 

The study of QCD within the systematic vertex expansion with full momentum dependences has been initiated for quenched \cite{Mitter:2014wpa} and unquenched \cite{Cyrol:2017ewj} vacuum QCD including dynamical hadronisation. Dedicated studies in Yang-Mills theory at finite and vanishing temperature, as well as in three dimensions in \cite{Cyrol:2016tym, Cyrol:2017qkl, Corell:2018yil}, are concentrating on the confinement aspects and its systematic. The respective results have also been used in a mixed scheme application to the full phase structure of QCD in \cite{Fu:2022gou} and the magnetic equation of state in \cite{Braun:2020ada}. Its systematic improvement in terms of higher vertices with full momentum dependences is work in progress in the fQCD collaboration \cite{fQCD}, aiming at apparent convergence of the scheme. 

This leaves us with an equally important remaining task: one has to control multi-scattering effects of fundamental and emergent degrees of freedom. In particular for QCD at high density with potentially competing order effects, first order regimes and mixed phases, the very efficient vertex expansion scheme with dynamical hadronisation has to be augmented with elaborate numerical methods for resolving the (multi-dimensional) effective potential, as well as the complicated momentum structure of scattering processes in a dense medium. An important recent technical advance has been the introduction and adaptation of state-of-the art numerical methods for the solution of general diffusion equations for the computation of the RG-flow of effective potentials \cite{Grossi:2019urj}, for subsequent works see \cite{Grossi:2021ksl, Koenigstein:2021rxj, Koenigstein:2021syz, Steil:2021cbu}. In \cite{Grossi:2021ksl} it has then been shown, that already the RG-flows for low energy effective models for high density QCD led to very steep structures in the effective meson potential and even shock-waves at high densities and low temperatures, that cannot be resolved by other methods commonly used in the field. In the present work we make use of the \textit{dune-FRGDG} framework put forward in \cite{Ihssen:2022xkr}, which uses Discontinuous Galerkin methods to solve the system of PDEs. 

We conclude our brief discussion of systematic expansions and their convergence in QCD with the remark, that the respective structural results and arguments straightforwardly carry over to general theories.

\subsection{Flowing fields at work}
\label{sec:FlowingPhi}

In the combined expansion scheme described in \Cref{sec:SysExpandQCD}, the momentum dependence of scatterings is taken into account within the vertex expansion. This expansion has been shown to converge rapidly if all the resonant channels are treated within dynamical hadronisation: the phase space suppression is very effective and in the absence of a small parameter this is called apparent convergence. In turn, multi-scattering events get important in the presence of close massless modes. Prominent examples in QCD are the pion in the chiral limit or the $\sigma$ and density modes close a potential critical end point. The slowing down of the convergence in the chiral limit has been studied in \cite{Braun:2020ada}: while for physical pion masses, $m_\pi \approx 140$\,MeV a Taylor expansion of the meson potential $V_\textrm{eff}(\rho_\varphi)$ with 
\begin{align}
\rho_\varphi=\frac12 \left(\sigma^2+ \bm \pi^2\right)\,, \quad \textrm{with} \quad \varphi^T=(\sigma,\pi_1,\pi_2,\pi_3)\,,
\end{align}
converges rapidly within 5 - 7 orders of $\rho_\varphi$, see also \cite{Mitter:2014wpa, Braun:2014ata, Cyrol:2017ewj, Fu:2019hdw}, far more orders are required for pion masses $m_\pi \lesssim 1$\,MeV, where also critical scaling sets in. The critical part of QCD in this regime is the mesonic sector, which is simply described by a (non-local) O(4)-model after integrating out all other degrees of freedom. 

The pivotal correlation function in these regimes is the full two-point function $\Gamma^{(2)}[\varphi_c](p)$ for constant backgrounds $\varphi_c$. Space-time or rather momentum-dependent backgrounds are covered by the higher order correlation functions $\Gamma^{(n)}[\varphi_c](p_1,...,p_n)$. In the vertex expansion their importance drops rapidly due to phase space suppression. Moreover, evaluated on the constant solution of the equations of motion $\varphi^{\ }_\textrm{EoM}$ with 
\begin{align}
\left.	\frac{\delta \Gamma[\varphi]}{\delta \varphi}\right|_{\varphi=\varphi^{\ }_\textrm{EoM}} =0\,,
\end{align}
the two point function $\Gamma^{(2)}[\varphi_\textrm{EoM}](p)$ in the mixed expansion converges rapidly towards that of the ground state of the theory, the optimal expansion point. 

The full two point function for constant fields is obtained by taking the second $\pi_i$-derivative at vanishing $\boldsymbol{\pi}=\bm 0$. It reads 
\begin{align}
	\Gamma^{(2)}_{\pi_1\pi_1}[\varphi](p)=& Z_\varphi(\rho_\varphi,p)\Bigl[ p^2 + m^2_\varphi(\rho_\varphi)\Bigr] \,,
\label{eq:FullG2} 
\end{align}
with the pole mass $m_\varphi(\rho_\varphi)$ defined by the on-shell condition $\Gamma_{\pi_1\pi_1}^{(2)}[\varphi](p^2= -m_\varphi(\rho_\varphi)) =0$. The wave function $Z_\varphi(\rho_\varphi,p)$ in \labelcref{eq:FullG2} is the coefficient of the operator $(\partial_\mu\varphi)^2$ in the effective action and the evaluation of the derivatives at $\boldsymbol{\pi}=\bm 0$ eliminates terms proportional to $\varphi$-derivatives of $Z_\varphi$ and $m^2_\varphi(\rho_\varphi)$. 

\Cref{eq:FullG2} leaves us with the particular task of determining the field- and momentum-dependent wave function $Z_\varphi(\rho_\varphi,p)$. Importantly, if the latter is non-trivial, this hints at the field $\varphi$ not being the physical field that leads to a simple description of the ground state and hence the theory. Moreover, its resolution is then a non-trivial numerical task. This suggests to dynamically change the field with the momentum scale such that the dispersion \labelcref{eq:FullG2} takes a simple classical form, $\varphi\to \phi(p,\varphi)$, leading to the dispersion 
\begin{align}
	\Gamma^{(2)}_{\phi_i \phi_j}[\phi](p)= \left[ p^2 + m^2_\phi(\rho) \right]\delta^{ij}\,,
	\label{eq:FullG2Optimal} 
\end{align}
or variants of it. This transformation can be included in terms of an fRG flow, where the transformation $\varphi\to \phi(\varphi,p)$ can be implemented and monitored successively as a function of the infrared cutoff scale $k$ with $\varphi\to \phi_k(\varphi,p)$. However, instead of using $\varphi\to \phi_k(\varphi,p)$, one is using the successive field transformation from the RG-adapted field basis $\phi$ at a given scale $k$ to the RG-adapted field basis at the scale $k-\Delta k$, encoded in the flow $\dot \phi_k(\phi,p)$. Here, the argument of $\dot \phi$ is the flowing field $\phi$ at the scale $k$. 

We emphasise that within this transformation the fields $\phi$ are the mean fields in the effective action and carry no $k$-dependence. Their relation to the fundamental fields is solely carried by the $k$-dependent function $\dot \phi_k(\phi,p)$. We now choose an RG-adapted field basis with $Z_\phi=1$ for all cutoff scales, 
\begin{align}
	\dot\phi_k(\phi,k) \qquad \longrightarrow \qquad Z_{\phi,k}(\phi,p)\equiv 1\,, 
\label{eq:FlowingFields} 
\end{align}
for all $k$, see \cite{Baldazzi:2021ydj}. \Cref{eq:FlowingFields} entails the differential form of the transformation from the microscopic fundamental fields to emergent effective fields with a classical dispersion. It is a specific choice in the class \textit{(iii)} of the general transformation $\dot\phi(\phi)$ in \labelcref{eq:FlowingFieldsGen}. 

A first application of \labelcref{eq:GenFlow} in the class \textit{(i)} to scalar O(N) theories can be found in \cite{Lamprecht2007}, where the reparametrisation of the theory in terms of flowing fields has been used to flow the fundamental fields $\varphi^T=(\varphi_1,...,\varphi_N)$ in a Cartesian basis into polar coordinates 
\begin{align}
	\phi^T= (\rho, {\bm \theta})\,,\quad \varphi=\sqrt{2\rho}\, e^{ \imag \theta^a t^a } \left(\begin{array}{c} 1\\ 0 \\[-1ex] \vdots \\ 0 \end{array} \right) \,,
\label{eq:PolarBasis}
\end{align}
Here, the $t^a$ with $a=1,...,N-1$ are the generators of the quotient O(N)/O(N-1), and the subgroup O(N-1) leaves the vacuum vector $(1,0,...,0)$ invariant. The field $\rho=\varphi^2/2$ in \labelcref{eq:PolarBasis} is the radial field, and the phase fields $\theta^a$ are the Goldstone fields. In the broken phase, the basis \labelcref{eq:PolarBasis} is advantageous, because the phase fields describe the Goldstone fields or rather their fluctuations for all values of the field, while in the Cartesian basis this only holds true at the expansion point $\varphi^T=(\varphi_1=\sigma,\bm 0)$. In turn, the polar basis has a parametrisation singularity for $\rho\to 0$ and hence in the symmetric phase the Cartesian basis is more natural. Accordingly, it is suggestive to define a basis that interpolates between the Cartesian basis in the symmetric phase (or small field values $\rho$) and the polar basis in the broken phase (or large field values of $\rho$). This is linked to an expansion scheme of the theory about the ground state of the theory, and has been called Goldstonisation in \cite{Lamprecht2007}. Notably, in the large N limit, the flow equation of the effective potential only depends on the Goldstone wave function $Z_\theta$. The compatibility of the results within the interpolating flowing fields with that in the standard Cartesian basis has been explicitly shown in \cite{Lamprecht2007}. 

The setup, notation and results have been also used in \cite{Isaule:2018mxt, Isaule:2019pcm} for an application to the thermodynamics of a Bose gases, for a more recent linear application see \cite{Daviet:2021whj}. There are many more applications including non-relativistic theories in and out of equilibrium, e.g.~ \cite{Mikheev:2018adp}, where a formulation in phase fields is better adapted to the physics phenomena at play, as one is setting up an expansion in the (dynamical) physical degrees of freedom. 

Such an application is complimentary to the one in the present work: while defining flowing fields that are adapted to the different phases of the theory, the formulation in \cite{Lamprecht2007, Isaule:2018mxt, Isaule:2019pcm, Daviet:2021whj} does not aim for a classical dispersion. To the contrary, in the large $N$ limit the whole effective potential is generated from the field-dependent wave function $Z_\pi(\rho,p)$ of the Goldstone modes.  

A more recent and intriguing application, directly aiming at \labelcref{eq:FullG2Optimal}, has been put forward in terms of the essential renormalisation group \cite{Baldazzi:2021ydj, Baldazzi:2021orb, Knorr:2022ilz, Knorr:2023usb}. In these works the essential RG has been used for the computation of critical scaling exponents and a stability analysis of fixed points both in scalar models and quantum gravity. In particular, the choice \labelcref{eq:FlowingFields} has been suggested in \cite{Baldazzi:2021ydj} and used in the three-dimensional O(1) model for a refined analysis of the Wilson-Fischer fixed point. In terms of the essential renormalisation group put forward in \cite{Baldazzi:2021ydj} the wave function $Z_\phi$ is an inessential coupling that can be absorbed in a redefinition of the field. As discussed above in the context of the O(N) application in \cite{Lamprecht2007}, in the large N limit the essential effective potential originates from the inessential wave function alone. This interesting structure and potential simplifications ask for a more detailed analysis. 

We emphasise that the current application as well as those in \cite{Lamprecht2007, Isaule:2018mxt, Isaule:2019pcm, Daviet:2021whj} are guided by the search for the optimal expansion and parametrisation of the theory about and in terms of the ground state, see also \cite{Salmhofer:2006pn}. This expansion about the full propagator, or covariance of the theory, is also is at the core of the RG-adapted expansion of the Wilson effective action and the 1PI effective action put forward in \cite{Ihssen:2022xjv} for fRG flows for complex actions. The respective functional optimisation setting is discussed in \cite{Pawlowski:2005xe}, and is related but not identical to a distinction of essential and inessential couplings. For a recent perspective of such a functional optimisation in terms of optimal transport \cite{Cotler:2022fze}. 

The above considerations and in particular the relation of optimal expansions to that about the ground state or full covariance of the theory also emphasise the importance of physical constraints for this general reparametrisation setup. For instance, instead of the form \labelcref{eq:FullG2Optimal} of the reparametrised kinetic operator or covariance obtained from $Z_\phi\equiv 1$ but unchanged mass function (in terms of the new field variable $\phi$), we may have chosen \labelcref{eq:FullG2Optimal} with $m^2_\phi(\rho)=0$ or any other mass function $m^2_\phi(\rho) \geq 0$. While such transformations exist and seemingly remove a relevant or marginal parameters (in $d\geq 2$) from the theory, they encode the expansion of a theory with pole mass $m_\phi^2(\rho)$ about theories with the pole mass $m_\phi^2(\rho)=0$ or any other choice. Within the present Euclidean setting this is a smooth transformation. However, while possible, it is certainly not optimal and functional optimisation in \cite{Pawlowski:2005xe} for the full two-point function shows this manifestly. Moreover, in the real-time or Minkowski version of the O(N) model, the respective transformation maps the spectrum of a given theory to a different one. Such a transformation necessarily moves pole and cut positions of correlation functions and has to be taken with a grain of salt. In any case it is hardly optimal. This concludes our brief discussion of general reparametrisations and flowing fields.

\section{Flowing fields for scalar O(N) theories }
\label{sec:GenRG}

In the following we explore the thermal phase transition in a scalar O(N) theory with the scalar field $\varphi^T=(\varphi_1,...,\varphi_N)$. This model has a thermal phase transition from a symmetric phase at high temperatures $T>T_c$ to a spontaneously broken phase for low temperatures $T<T_c$ with one radial mode and $N-1$ massless Goldstone modes. 

As discussed in \Cref{sec:FlowingPhi}, we use the reparametrisation freedom stored in the flowing fields \labelcref{eq:FlowingFields} to transform the full dispersion into a classical one for all field values, while keeping the pole mass fixed. To that end we consider the following approximation of the effective action in the presence of an infrared cutoff scale, 
\begin{align}
	\hat \Gamma_k[\varphi] = \int_x\, \Big[ \frac{1}{2} Z_{\varphi}(\rho_\varphi) (\partial_\mu \varphi)^2 + V_{k}(\rho_\varphi) \Big] \,,
	\label{eq:effActionhat}
\end{align}
with the notation 
\begin{align}
	\int_x = \int_0^\beta dx_0 \int_{\mathbbm{R}^3} d^3 x\,, \qquad \rho_\varphi= \frac{\varphi^2}{2}= \frac{\varphi^a \varphi^a }{2}\,,
\end{align}
where $\beta= 1/T$ with the temperature $T$, and $a=1,...,N$. \Cref{eq:effActionhat} corresponds to the first order of the derivative expansion for the O(N) model. 

For $N=4$ this model is a simple low energy effective theory for the chiral dynamics in QCD, and the phase transition is that of strong chiral symmetry breaking, as also discussed in the last section. The radial scalar $\sigma$-mode is obtained from the quark bilinear $\bar q q$. The three (pseudo-)Goldstone bosons are related to the three pseudoscalar pions ${\bm \pi}$ obtained from $\bar q \gamma_5 \boldsymbol{\sigma}q$ with the Pauli matrices $\boldsymbol{\sigma}=(\sigma^1, \sigma^2, \sigma^3)$. The masses $m_\pi \approx 140$\,MeV of the latter are generated by an explicit symmetry breaking term linear in the radial field. This simple model emerges dynamically as part of the full QCD effective action within the fRG approach to first principle QCD with dynamical hadronisation, for recent works see~\cite{Fu:2019hdw, Fukushima:2021ctq} and the reviews \cite{Dupuis:2020fhh, Fu:2022gou}. While we will use the results of this specific application in a forthcoming QCD related work, the approach and results here are far more general. 

Now we use a transformation \labelcref{eq:FlowingFields}, that maps the effective action $\hat \Gamma_k[\varphi]$ to one with a classical dispersion in terms of the transformed field $\phi$ with 
\begin{align}
\Gamma_k[\phi] = \int_x \Big[ \frac{1}{2} (\partial_\mu \phi)^2 + V_{k}(\rho) - c_\sigma \sigma \Big] \,, 
	\label{eq:effAction}
\end{align}
where 
\begin{align} 
	\phi=\left( \begin{array}{c}\sigma\\ \boldsymbol{\pi}\end{array}\right) \,,\qquad \rho = \frac{\phi^2}{2}= \frac{\sigma^2 +\boldsymbol{\pi}^2}{2}\,.
\end{align}
and $\boldsymbol{\pi}^T=(\pi_1,...,\pi_{N-1})$. In \labelcref{eq:effAction} we have introduced an explicit linear breaking term for the radial mode $\sigma$. Note that such a term drops out of the flow and only shifts the (constant) solution $\phi_{\mathrm{EoM}}$ of the equation of motion, 
\begin{align}\label{eq:EoM}
	\sigma \partial_{{\rho}} V_k( \rho_0) = c_\sigma \,.
\end{align}
Accordingly, this linear term is only a spectator in the following derivations and computations. 
We emphasise that the form of \labelcref{eq:effAction} does not signal an approximation with the assumption $Z_\phi\approx 1$, but a formulation with flowing fields with the constraint 
\begin{align}
	Z_\phi(\rho) \stackrel{!}{\equiv} 1\,, 
\label{eq:Z=1}
\end{align}
which is the momentum-independent reduction of \labelcref{eq:FlowingFields}. 

We also envisage \labelcref{eq:effAction} with \labelcref{eq:Z=1} as the intermediate reparametrisation of the theory, before the Goldstonisation of \cite{Lamprecht2007, Isaule:2018mxt, Isaule:2019pcm, Daviet:2021whj} is applied. This combined application is deferred to future work.

\subsection{RG-adapted field transformations}\label{sec:RG-kernel}

It is left to derive the flowing fields $\dot{\phi}$ in \labelcref{eq:FlowingFields}, that absorbs the full field dependence of $Z_{\phi,k}$ into the field $\phi$ and leads to 
\begin{align}
	\partial_t Z_\phi(\rho) =0\,, 
\label{eq:dtZ=0}
\end{align}
where we have dropped the subscript ${}_k$ indicating the $k$-dependence. To begin with, both actions, \labelcref{eq:effActionhat,eq:effAction} are invariant under linear O(N) transformations, so we only have to consider field transformations that carry a linear representation of O(N). Consequently, a general parametrisation of $\dot \phi$ is given by 
\begin{align}
	\dot\phi = - \frac12 \eta_\phi(\rho) \, \phi \,,
\label{eq:dotphieta}
\end{align}
with a general $\rho$-dependent function $\eta_\phi(\rho)$, which is indeed absorbing the anomalous dimension of the field, hence the notation. 

In order to determine \labelcref{eq:dotphieta}, we compute the flow of $Z_\phi$ from that of the two-point function, $\partial_t \Gamma^{(2)}_k[\phi](p)$, the second field derivative of the functional flow with flowing fields, \labelcref{eq:GenFlow}. As discussed below \labelcref{eq:FullG2}, the wave function is directly extracted from the two-point function of the Goldstone fields at $\boldsymbol{\pi}=\bm 0$. The general form of the Goldstone two-point function for constant fields $\phi=(\sigma, \bm 0)$ and $\rho =\sigma^2/2$ is given by 
\begin{align}
\Gamma_{k,\pi_i\pi_j}^{(2)}[\phi](p) = Z_\phi(\rho,p)\,\left[p^2 +m_\phi(\rho)\right] \delta^{ij} \,. 
\label{eq:GenG2}
\end{align}
It is diagonal in field space and the respective term in the effective action is given by 
\begin{align}\nonumber 
\Gamma_k[\phi] = & \frac12  \int_x \Biggl[ (\partial_\mu \phi_n) Z_\phi(\rho,p)\, (\partial_\mu \phi_n)\\[1ex]
& \hspace{1cm}+ \phi_n m_\phi(\rho) Z_\phi(\rho,p) \phi_n\Biggr]+\cdots \,, 
	\label{eq:GenG}
\end{align}
where $\cdots$ stands for the interaction terms and $p=\imag \partial$. \Cref{eq:GenG} entails that the $\pi_i, \pi_j$-derivatives leading to \labelcref{eq:GenG2} have to hit both first derivative terms  $\partial_\mu \phi_n$ as $\partial_\mu \phi_n=0$ for constant fields. The mass parameter $m_\phi(\rho )$ is the pole mass of the theory as discussed below \labelcref{eq:FullG2}. While the pole mass can be accessed within the fRG approach, in particular within spectral flows, see e.g.~\cite{Braun:2022mgx}, we will use an additional approximation in the present work with 
\begin{align}
Z_\phi(\rho,p)\approx Z_\phi(\rho)\,.
\label{eq:MomIndepZ}
\end{align}
With \labelcref{eq:MomIndepZ} the flow of $Z_\phi(\rho)$ is computed by taking a $\boldsymbol{p}^2$ derivative of $\partial_t \Gamma_k^{(2)}[\phi](p)$ at $p=0$, as all mass terms drop from the flow of \labelcref{eq:GenG2}. In combination we arrive at 
\begin{align}
	\partial_t Z_\phi(\rho) =\left. \frac{\partial}{\partial \boldsymbol{p}^2} \partial_t \Gamma^{(2)}_{k,\pi_1 \pi_1}[\phi](p)\right|_{\small{\begin{array} {rl} p=&\!\!0\\ \boldsymbol{\pi}=&\!\! \bm 0\end{array}}} \,.
\label{eq:Zflow}
\end{align}
The details of the explicit computation are deferred to \Cref{app:Zprojection}, and the final flow for $p=0$ and $\boldsymbol{\pi}= \bm 0$ reads 
\begin{align}\nonumber 
	\partial_t Z_\phi=&\, \frac12 \frac{\partial}{\partial \boldsymbol{p}^2} \, \Tr \left[G^{(2)}_{,\pi_1 \pi_1}\left(\partial_t + 2\frac{\delta \dot{\phi}}{\delta \phi} \right) R_k \right]	\\[1ex] 
	&- 2 Z_\phi \frac{\partial  \dot{\pi}_1 }{\partial \pi_1} +2  \rho Z'_\phi \eta_\phi \,,
\label{eq:Zflow2}
\end{align}
where $Z'_\phi=\partial_\rho Z_\phi$ and$G^{(2)}_{\pi^1 \pi^1}$ stands for the second derivative of the propagator w.r.t.~the Goldstone $\pi_1$, and we have in general
\begin{align}\nonumber 
	G^{(m)}_{\phi_i\phi_j,\phi_{n_1}\cdots \phi_{n_m}}(p_1,...,p_{m+2})& \\ 
	&\hspace{-1cm} = \frac{G_{\phi_i\phi_j}(p_1,p_2)}{\delta \phi_n(p_3)\cdots \delta\phi_m(p_{m+2})}\,.
\label{eq:G2Notation}
\end{align}
Similarly to the short-hand notation introduced below \labelcref{eq:GnNotation} for $\Gamma^{(n)}$ and below \labelcref{eq:Prop} for the propagator, we shall also use $G_{\phi_i\phi_j,\phi_n\phi_m}$, $G_{i j,n m}$ as well as $G_{,\phi_n\phi_m}$ and extensions to more derivatives for the sake of readability. 

Again using \labelcref{eq:dotphieta} and $\pi_i=0$, the last term in \labelcref{eq:Zflow2} reduces to 
\begin{align}
- 2 Z_\phi \frac{\partial  \dot{\pi}_1 }{\partial \pi_1} +2  \rho Z'_\phi \eta_\phi = Z_\phi \eta_\phi +2  \rho Z'_\phi \eta_\phi \,.
	\label{eq:Z+rhoZ'Reduce}
\end{align}
Using $Z_\phi=1$, due to the constraint imposed in \labelcref{eq:Z=1}, at any given $k$, all derivatives of $Z_\phi$ in the second line of \labelcref{eq:Zflow2} vanish. Moreover, for the choice 
\begin{align}
 Z_\phi \eta_\phi +2  \rho Z'_\phi \eta_\phi =\frac{1}{2} \frac{\partial}{\partial \boldsymbol{p}^2} \Tr \left[G^{(2)}_{,\pi_1 \pi_1}\left(\partial_t + 2\frac{\delta \dot{\phi}}{\delta \phi} \right) R_k \right] \,, 
\label{eq:dynamHadron}
\end{align}
for $p=0$ and $\boldsymbol{\pi}= \bm 0$ we obtain $\partial_t Z_\phi=0$ and hence \labelcref{eq:Z=1} is sustained for all $k$. In \labelcref{eq:dynamHadron}, we have used the short hand notation $G^{(2)}_{,\pi_1 \pi_1}$, where we have dropped the indices that are involved in the trace. 

Comparing \labelcref{eq:dynamHadron} with the $\pi_1$-derivative of \labelcref{eq:dotphieta} at $p=0$ and $\boldsymbol{\pi}=\bm 0$ leads us to 
\begin{align}
\eta_\phi  =- \frac{1}{2} \frac{\partial}{\partial \boldsymbol{p}^2} \Tr \left[G^{(2)}_{,\pi_1 \pi_1}\Bigr(\partial_t - \eta_\phi - 2 \rho \, \eta_\phi'\Bigl) R_k \right] \,,
	\label{eq:dynamHadron-eta}
\end{align}
where $\eta_\phi'=\partial_{{\rho}} \eta_\phi$. It is worth noting that \labelcref{eq:dynamHadron-eta} with $\eta_\phi'=0$ is also obtained, if $R_k$ in the standard fRG approach without flowing fields is augmented with a field-dependent wave function, 
\begin{align}
	R_k(\rho,p) = Z_\phi(\rho) R_k^{(0)}(p)\,. 
\label{eq:BackR}
\end{align}
\begin{figure*}
	\centering
	\begin{subfigure}{.48\linewidth}
		\centering
		\includegraphics[width=0.95\linewidth]{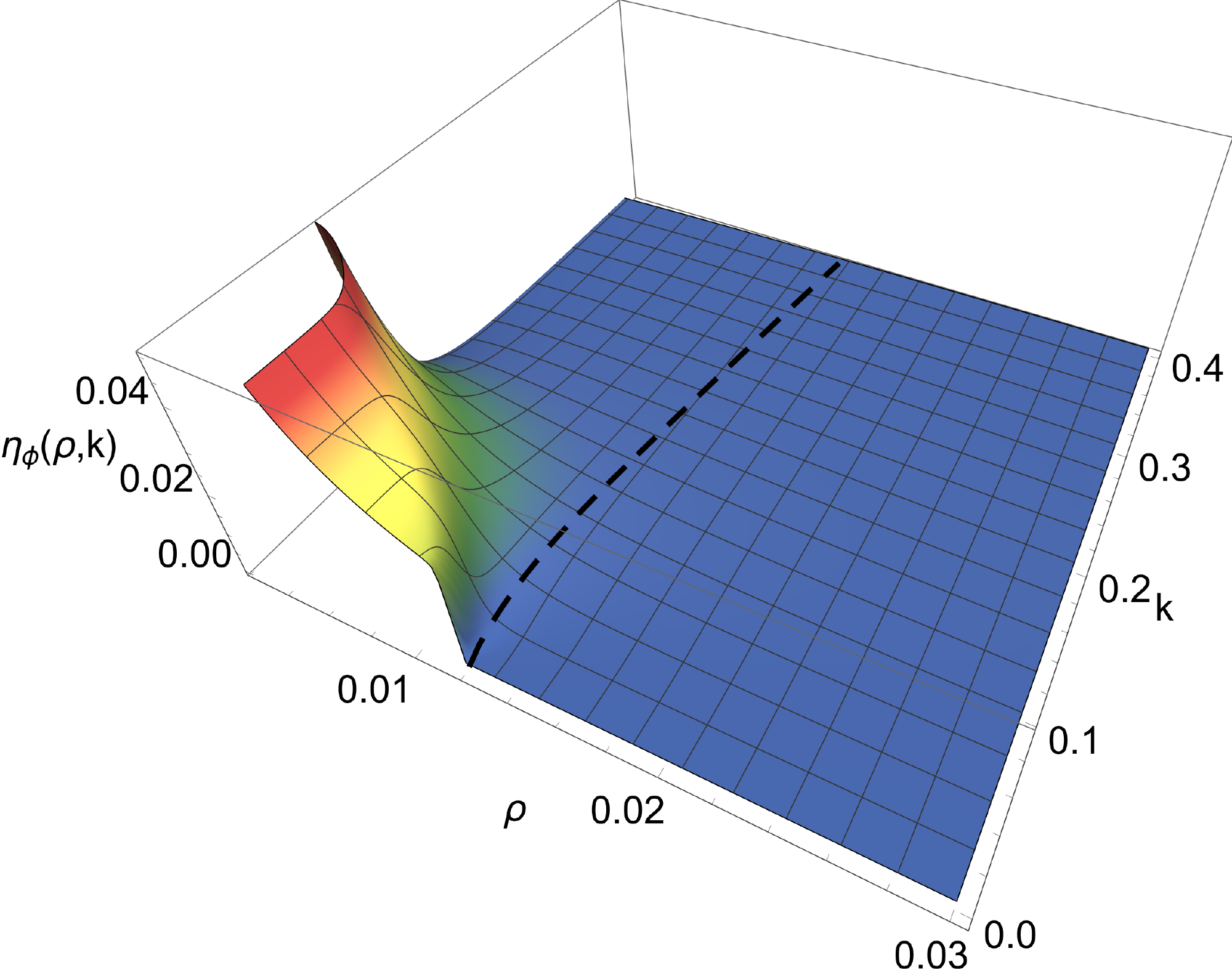}
		\caption{Anomalous dimension in the broken phase for $T=0.100$ deep in the broken phase.\hspace*{\fill}} \label{fig:dotPhiinVac}
	\end{subfigure}%
	\hspace{0.35cm}%
	\begin{subfigure}{.48\linewidth}
		\centering
		\includegraphics[width=0.95\linewidth]{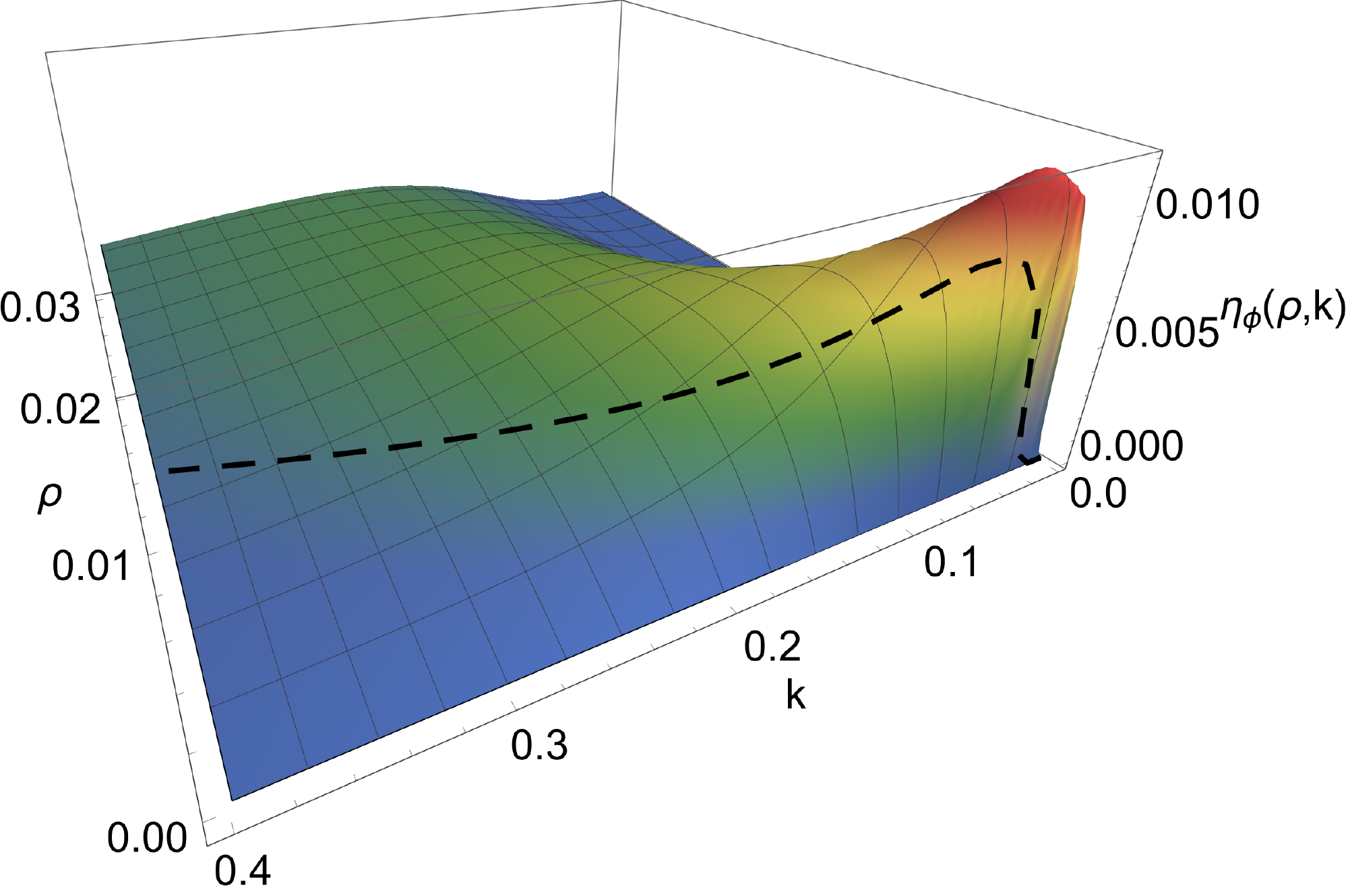}
		\caption{Anomalous dimension in the symmetric phase for $T=0.262$ close to the critical temperature.\hspace*{\fill}} \label{fig:dotPhiathighT}
	\end{subfigure}
	\caption{Field and RG-time dependence of the anomalous dimension in the broken phase, \Cref{fig:dotPhiinVac}, and in the symmetric phase, \Cref{fig:dotPhiathighT}, in four dimensions at finite temperature. The critical temperature is $T_c \approx 0.260$ and the initial conditions are given in \Cref{sec:resd4}. All units are given in terms of the UV-cutoff $\Lambda = 1$, and the dashed black line indicates the equations of motion in the limit of massless Goldstone bosons. \hspace*{\fill}}
	\label{fig:dotPhi}
\end{figure*}
\Cref{eq:BackR} is at the root of the background approximation in scalar theories, gauge theories and gravity. However, \labelcref{eq:BackR} leads to higher loop terms in the flow equation. In the background approximation the higher loop terms are simply dropped. Then we are led to \labelcref{eq:dynamHadron-eta} as well as the flow \labelcref{eq:GenFlow} with \labelcref{eq:dotphieta}, where we consistently approximated $\eta_\phi'\approx 0$. This holds true in the limit of slowly varying $\eta_\phi$, which is analogous to the local density approximation. Note that while the validity regime of this approximation is difficult to estimate, as the size of $\eta_\phi'$ is not the relevant quantity, but rather its impact if fed back to the flow. The latter impact can be estimated in the fRG within a (linear) self-consistency analysis: we can feed back the $\eta_\phi'$ obtained in the $\eta_\phi'\approx 0$ approximation into the right hand side of the flow and use the difference as an error estimate. 

Consequently, the fRG approach with flowing fields supports the background field approximation within the approximation \labelcref{eq:MomIndepZ} in the limit of slowly varying $\eta_\phi$. Importantly, within the approach with flowing fields, this approximation can be systematically lifted without leading to higher order loop terms. In particular, momentum-dependent $Z_\phi(\rho,p)$ can be implemented straightforwardly. 

The full RG-adapted reparametrisation of the theory in terms of a non-linear transformation in field space \labelcref{eq:dynamHadron} is linked to linear ones, if further approximations are used. Already in the derivation of \labelcref{eq:dynamHadron} we have dropped the momentum dependence of the wave function. While not necessary, it simplifies the final transformation \labelcref{eq:dynamHadron} significantly. A further significant simplification is achieved if we drop the field dependence in \labelcref{eq:dynamHadron}, which linearises the RG-adapted transformation and reduces it to a cutoff-dependent rescaling of the field. This approximation is a commonly used one, the local potential approximation (LPA) with cutoff-dependent but field-independent wave functions, called LPA$'$. In contrast, the full RG-adapted fields with \labelcref{eq:dynamHadron} encode the full first order of the derivative expansion. However, the flows in the latter depend on derivatives of $Z_\phi(\rho)$ which are absent here, due to the RG-adaptation. 

The LPA$'$ approximation is obtained by \labelcref{eq:effActionhat} with $Z_\varphi(\rho_\varphi) \to Z_\varphi(\rho_0)$ with a specific $\rho_0$, 
\begin{align}
	\Gamma^{\textrm{\tiny{LPA$'$}}}_k[\varphi] = \int_x\, \Big[ \frac{1}{2} Z_{\varphi}(\rho_{0}) (\partial_\mu \varphi)^2 + V_{k}(\rho_\varphi) - c_\sigma \sigma \Big] \,. 
	\label{eq:LPA$'$}
\end{align}
Typically, $\rho_{0, k}$ is chosen as the cutoff-dependent solution of the equations of motion \labelcref{eq:EoM}. The rescaling of the field is then given by,
\begin{align}\label{eq:anomalousDim}
	\dot{\phi} = - \phi \frac{ \eta_\phi(\rho_{0})}{2}\,, \qquad \qquad \eta_\phi(\rho_{0,k})= -\frac{\partial_t Z_\varphi (\rho_{0,k})}{Z_\varphi (\rho_{0,k})} \,.
\end{align}
In short, in LPA$'$ the flowing fields leading to \labelcref{eq:Z=1} are simply implementing the rescaling $\phi= Z^{(1/2)}_\varphi(\rho_0) \varphi$. Conceptually, the difference between the full field-dependent reparametrisation and the LPA$'$ one is the additional assumption, that the field dependence of $Z_\varphi$ is negligible, $\partial_{\rho_\varphi} Z_{\varphi} \approx0$. This is the analogue of a local density approximation in density functional theory. This relation of the fRG flows with flowing fields to the well-studied LPA$'$ approximation allows us to gauge the impact of fully field-dependent reparametrisations. Consequently, in \Cref{sec:results} we shall compare the full results with that obtained in LPA$'$. 

\subsection{Flow equations}
\label{sec:PotFlow}

In the present approximation we are left with two flow equations, one for the effective potential and that for the flowing field. 
With the underlying O(N) symmetry we can formulate the flows in terms of $\rho$ instead of $\phi$. Here we provide the explicit forms of the two flows in the present setting of a thermal O(N) model. In comparison to the flow of the effective potential in LPA, the flow carries additional terms proportional to $\eta_\phi(\rho)$ and its $\rho$-derivative, 
\begin{align}\nonumber 
	\partial_t V_k -\eta_\phi \rho V'= &\,  k^{d+1} A_d \Bigg[ \!\left(1 - \frac{\eta_\phi + 2 \rho \, \eta_\phi'}{1+d} \right)\frac{\coth \left(\frac{\epsilon_\sigma}{2T} \right)}{2 \, \epsilon_\sigma} \\[1ex]
	 &\hspace{-.5cm}+ \left(1 - \frac{\eta_\phi}{1+d}\right)\frac{(N-1) \coth \left( \frac{\epsilon_\pi}{2T} \right)}{2 \, \epsilon_\pi}
	 \Bigg] 
	\,, 
\label{eq:flowV}
\end{align}
with $V'=\partial_\rho V$, $V''=\partial^2_\rho V$, $\eta_\phi'=\partial_\rho \eta_\phi$ and
\begin{align}
	A_{d+1} = \frac{2 \pi^{d/2} }{ (2 \pi)^{d}\Gamma(d/2) d} \,.
\end{align}
The dispersion relations $ \epsilon_{\sigma/\pi}$ are given by 
\begin{align}
\epsilon_{\pi/\sigma} = \sqrt{k^2 + m_{\pi/\sigma}^2}\,,
\label{eq:Dispersion}
\end{align}
and the radial and Goldstone masses 
\begin{align}
	m_\pi^2 = V'\,,\quad m_\sigma^2 &=m_\pi^2 + 2\rho V'' \,.
\label{eq:masses}
\end{align}
The RG-adapted flow of the field basis is given by \labelcref{eq:dynamHadron-eta} and has the explicit form in terms of $\eta_\phi$,
\begin{align}
	\eta_\phi = 4
		A_d\,\bar\rho (\bar V'')^2\left(1-\frac{ \eta_\phi + 2 \, \rho \eta_\phi'}{1+d} \right)  \mathcal{BB}_{(2,2)} \,,
\label{eq:Exdotphi1}
\end{align}
\begin{figure*}
	\centering
	\begin{subfigure}{.495\linewidth}
		\centering
		\includegraphics[width=\linewidth]{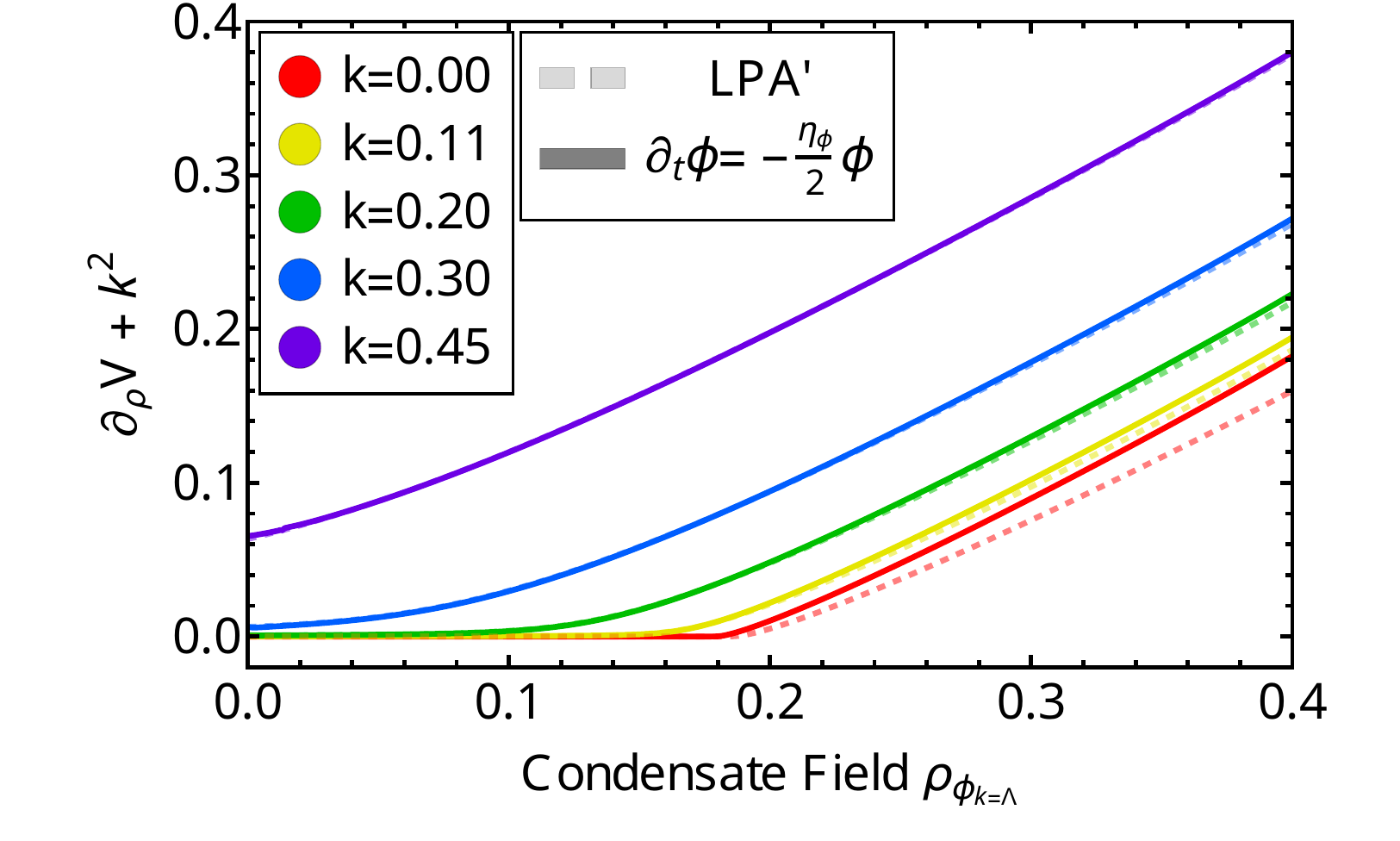}
		\caption{(1PI) First field derivative $\partial_\rho V(\rho)$ of the potential. \hspace*{\fill}} \label{fig:mpi3}
	\end{subfigure}%
	\hspace{0.01\linewidth}%
	\begin{subfigure}{.495\linewidth}
		\centering
		\includegraphics[width=\linewidth]{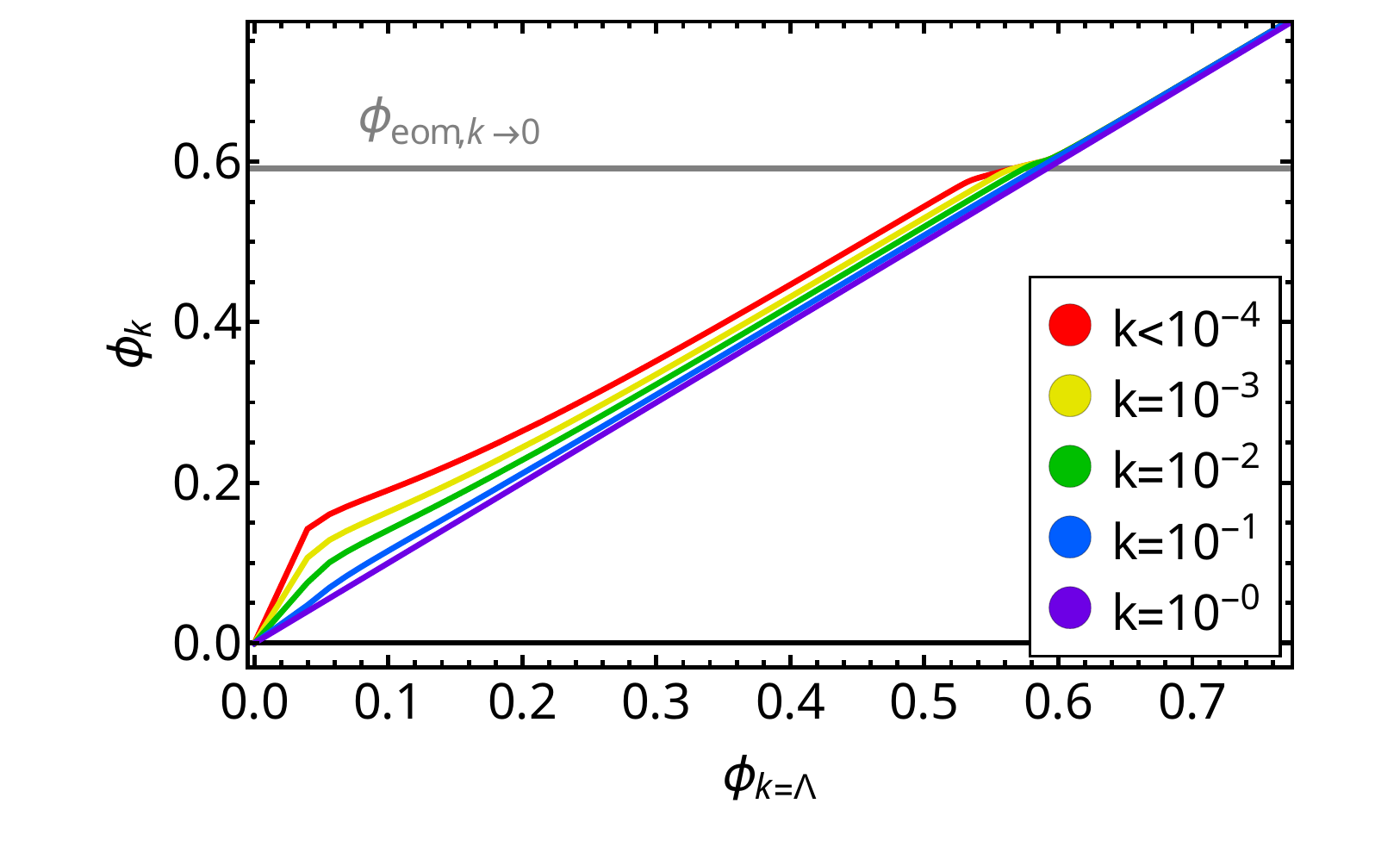}
		\caption{Flowing field $\phi_k(\varphi)$ as a function of the UV field $\varphi=\phi_{k=\Lambda}$.\hspace*{\fill}} \label{fig:varphi3}
	\end{subfigure}
	\caption{Comparison of the RG-scale dependence of the Goldstone mass $u = \partial_{{\rho}} V + k^2$ (a) and the flowing field $\phi_k$ (b). Results are shown in the broken phase in $d=3$ at $T=0.01$ for the initial conditions specified in \Cref{sec:resd3}. We compare results of the approximation with field-dependent $Z_\phi(\rho)$ to LPA$'$ (dashed lines) and indicate the field dependence in terms of $\rho_\varphi = \rho_\Lambda$. All units are given in terms of the UV-cutoff $\Lambda = 1$.\hspace*{\fill}}
	\label{fig:pot3}
\end{figure*}
with the dimensionless threshold function $\mathcal{BB}_{(2,2)}$ defined in \labelcref{eq:BB22} in \Cref{app:thrs}, and dimensionless fields $\bar\rho = k^{d-2} \rho$ and potential $\bar V = k^d \, V$. A solution for the anomalous dimension $\eta_{\phi,k}(\rho)$ for $d=4$ dimensions for temperatures $T>T_c$ in the symmetric and $T<T_c$ in broken phase is depicted in \Cref{fig:dotPhi}. Note that the the right hand side of \labelcref{eq:Exdotphi1} tends towards zero in the large N limit with N$\to\infty$. This is consistent with the vanishing anomalous dimension in this limit. 

We close this analysis with the comment that this setup can be used as a starting point for an additional transformation of the field basis into polar coordinates in the broken phase, with or without a consequent absorption of the emergent wave function $Z_\theta(\rho)$. Such a set-up is fully adapted to the ground state or covariance of the O(N) theory. 

\section{Results}
\label{sec:results}

We present results for an O(4) theory in $d=3$ dimensions in \Cref{sec:resd3}. Due to an easier numerical access, $d=3$ is the ideal testing ground to investigate the behaviour of the anomalous dimension for a set of initial conditions which starts deep in the broken phase. In \Cref{sec:resd4} we present numerical results for the thermal phase transition in an O(4) theory in $d=4$ dimensions, which emulates a QCD-like setting. This includes a systematic error estimate for the critical temperature, obtained in an LPA$'$ approximation that is commonly used within these models. For details on the numerical evaluation see \Cref{sec:finiteDiff}.

\begin{figure}
	\centering
	\includegraphics[width=\linewidth]{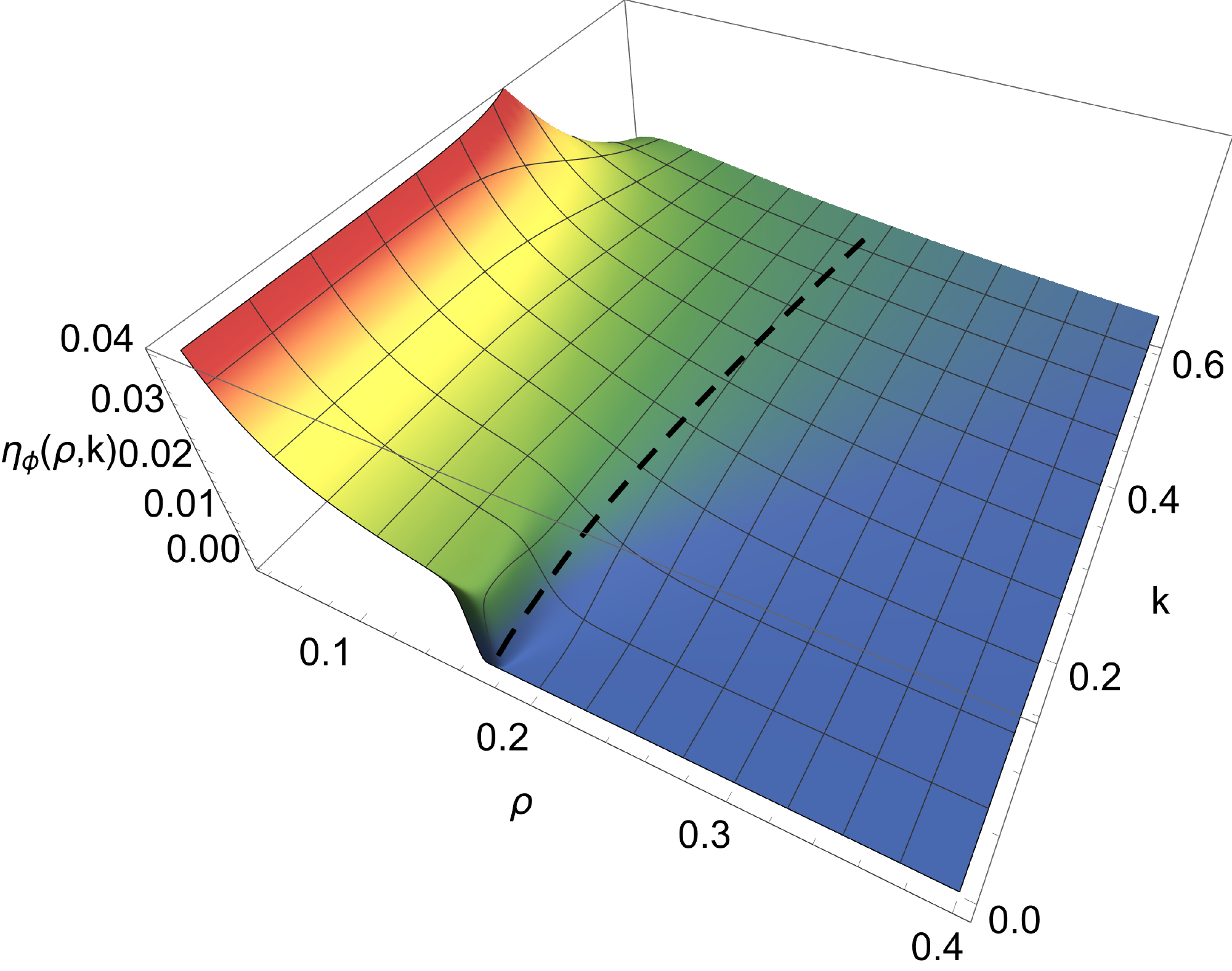}
	\caption{Anomalous dimension in $d=3$ deep in the broken phase ($T=0.01$). We find a strong field dependence in the unphysical (flat) part of the potential $\rho<\rho_0$, where the latter is the solution to the equations of motion (\labelcref{eq:EoM}). For orientation, the dashed black line indicates the solution to the equations of motion (\labelcref{eq:EoM}) in the limit of massless Goldstone bosons. All units are given in terms of the UV-cutoff $\Lambda = 1$.\hspace*{\fill}\hspace*{\fill}}
	\label{fig:dotPhiZoom}
\end{figure}
%
\subsection{Flowing fields in the broken phase}
\label{sec:resd3}

In this section we present results in $d=3$ dimensions, where the flow is initiated deep in the broken phase. The initial conditions of the RG-time integration are given by a $\phi^4$-potential at an initial RG-scale of $\Lambda = 1$, to wit
\begin{align}\label{eq:iniPot}
	u(\rho) = \partial_{{\rho}} V(\rho) = \rho \lambda + m^2 \,.
\end{align}
with $\lambda =1$ and $m^2 =- 0.25$. 

We depict the RG-scale dependence of the solution to the flow of the potential derivative (the Goldstone mass) \labelcref{eq:schemFlow} in \Cref{fig:mpi3}. This result is compared to an LPA$'$ computation with the same initial conditions. 
Both computations compare well on the solution of the equations of motion $\rho_0$, recall \labelcref{eq:EoM}. The relative deviation of $\rho_0$ between both calculations is $<1\%$ at all times during the RG-time evolution.
However, at some finite RG-scale $k>0$, the approximations start deviate away from the equations of motion in the pysical regime $\rho> \rho_0$.
Here, the deviations emphasise the improvement that comes with the flowing fields formulation, or rather its more complete field-dependences. It also translates to an improved evaluation of higher order couplings. 

In summary, the current result, in a qualitatively improved approximation, validates the use of the LPA$'$. Both methods agree on the EoM, where the physical correlation functions are obtained. This is not surprising, as the anomalous dimension on the EoM is relatively small and flat away from the flat regime, see \Cref{fig:dotPhiZoom}. We expect this to change for increasingly more dynamical settings, such as e.g. fermionic models with finite density. Note also, that the deviations away from the EoM indicate, that higher correlation functions may differ even on the EoM. 

The flowing field formalism introduces a $k$-dependence to the coordinate space via the relation \labelcref{eq:dotphieta}. The underlying transformation $\phi_k[\varphi]$ is obtained by integrating the flow $\dot\phi$, 
\begin{align}
	\phi_k = \phi_\Lambda +\int_{k}^{\Lambda} \frac{\mathrm{d} k }{2 k} \, \eta_{\phi,k} \, \phi_k \,, 
	\label{eq:dotphieta2}
\end{align}

with $\varphi=\phi_\Lambda$. In \Cref{fig:varphi3}, we show the integrated flowing fields as a function of the initial UV-fields. 
We can see a flattening of the coordinates around $\rho_0$. This coincides with the jump in the field dependence of the anomalous dimension $\eta(\rho)$ below $\rho_0$. We suspect that in even more elaborate momentum approximation schemes, this flattened region will increase in size. 
\begin{figure}
	\centering
	\includegraphics[width=\linewidth]{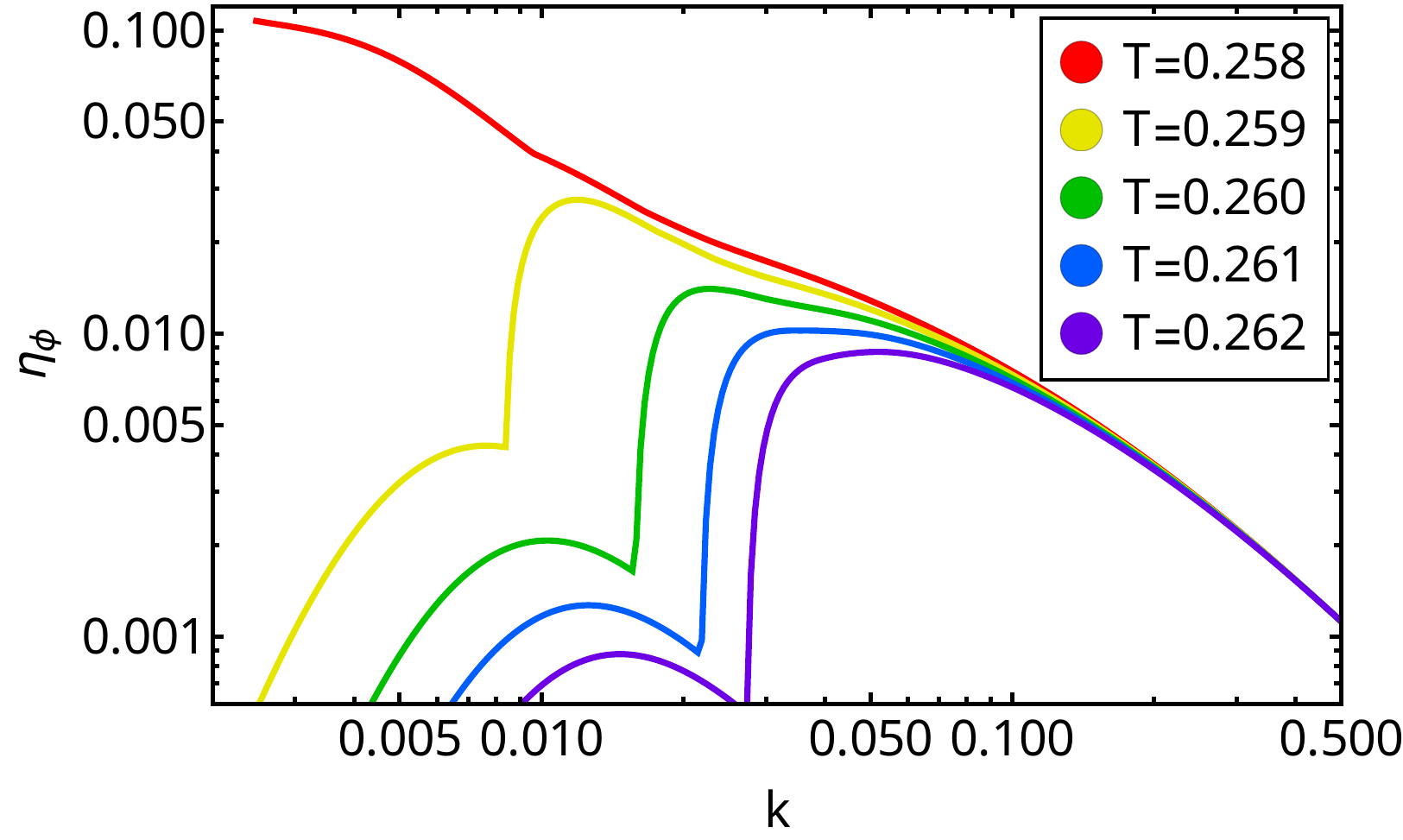}
	\caption{RG-scale dependence of the anomalous dimension $\eta_{\phi}(\rho_{0})$ on the equations of motion \labelcref{eq:EoM}, in the vicinity of the critical temperature $0.258<T_c<0.259$. The current result is compatible with a scaling solution $\eta_{\phi,c} \approx 0.042$. \hspace*{\fill}}
	\label{fig:etaCrit}
\end{figure}
%
\subsection{Thermal phase transition in $d=4$ dimensions }
\label{sec:resd4}

In this section we apply the flowing field approach to the thermal phase transition in the four-dimensional $\phi^4$ theory. The effective action at the initial scale $\Lambda=1$ is given by the classical potential \labelcref{eq:iniPot} with the coupling $\lambda= 1$ and the mass parameter $m^2 = -0.025$. All units are measured in terms of the UV-cutoff scale $\Lambda$. 

Note that this choice of the initial effective action as the classical one is not fully RG-consistent \cite{Braun:2018svj, Pawlowski:2005xe} due to the relatively large coupling. However, the second order thermal phase transition in this theory occurs at sufficiently low temperature $T_c \approx \Lambda/4$, and hence potential ultraviolet effects due to the choice of the classical action as the initial effective action have decoupled. Moreover, in low energy effective theories of QCD the initial $\phi^4$ coupling is typically even larger.

\subsubsection{Phase transition regime}
\label{sec:PT}
\begin{figure}
	\centering
	\includegraphics[width=\linewidth]{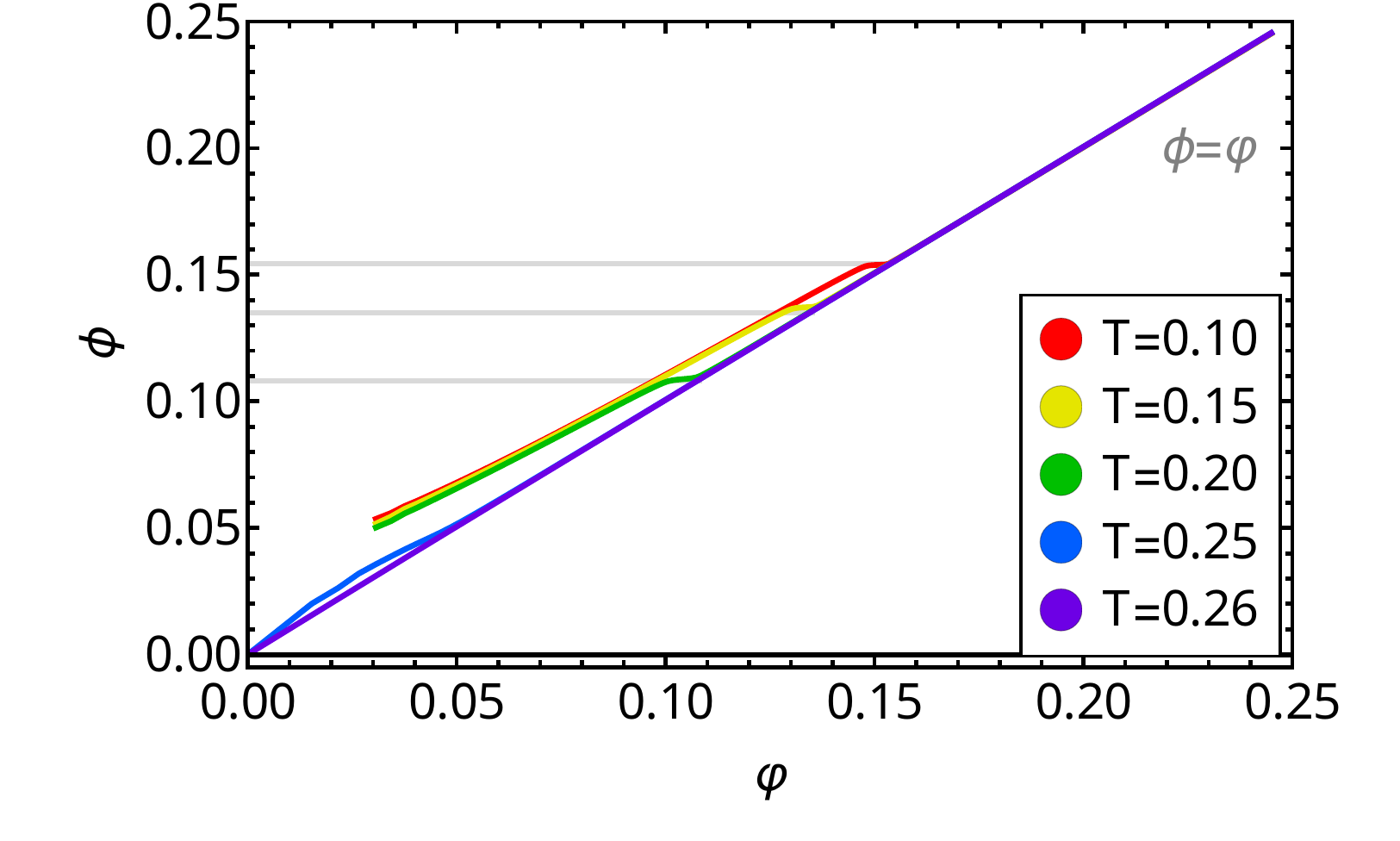}
	\caption{Temperature dependence of the dynamical fields as a function of the UV-field at $k=0.002$. All units are given in terms of the UV-cutoff $\Lambda = 1$. We indicate the prolongation of the flat part by grey lines, which correspond to the solution to the equations of motion. We do not indicate the small field values, since they are corrected in the numerical evaluation, see \Cref{sec:statEq}. \hspace*{\fill}}
	\label{fig:flowingT}
\end{figure}
\begin{figure*}
	\centering
	\begin{subfigure}{.495\linewidth}
		\centering
		\includegraphics[width=\linewidth]{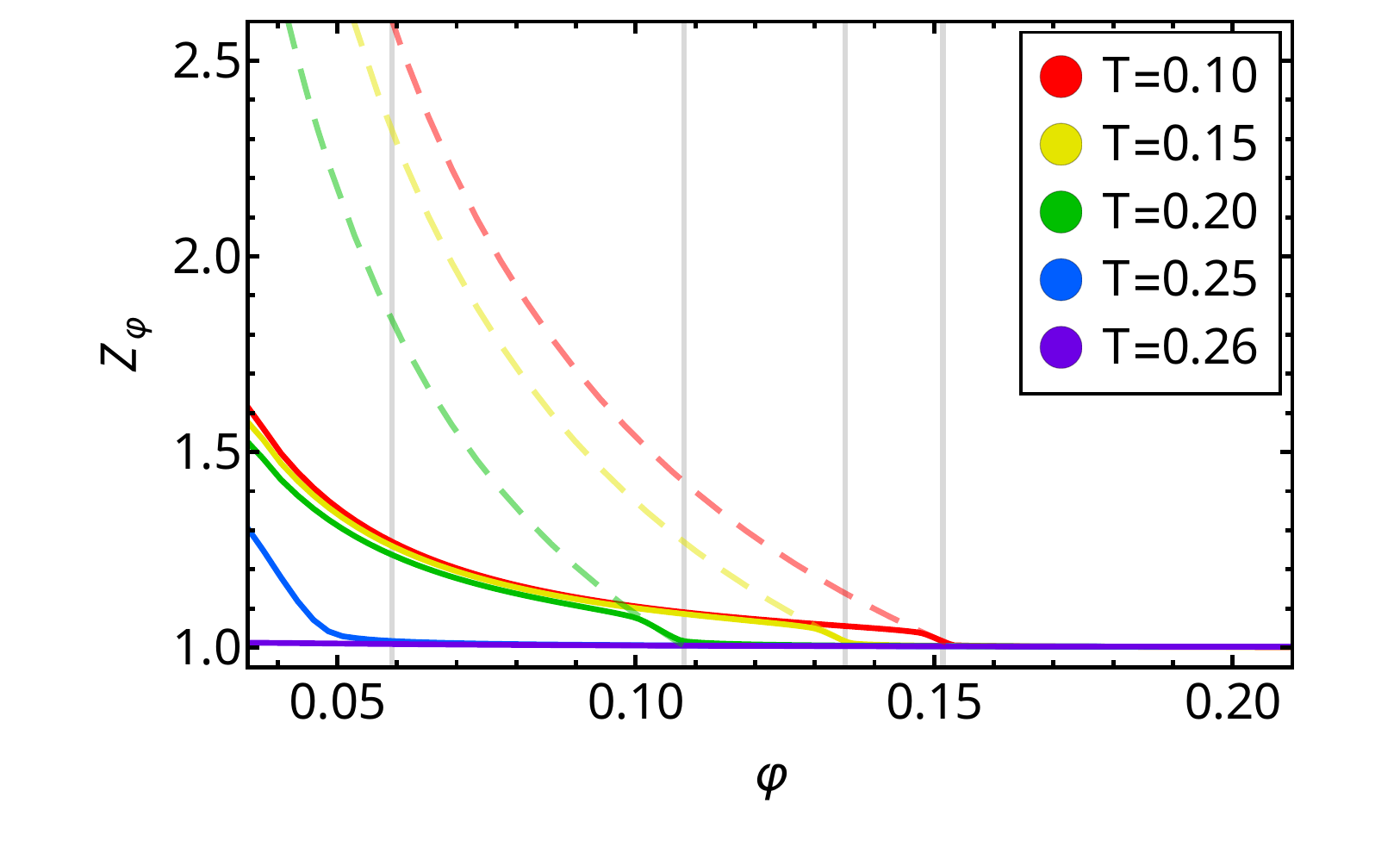}
		\caption{Wave function $Z_\varphi(\varphi)$ at $k = 0.002$. \hspace*{\fill}} \label{fig:fielddepZ}
	\end{subfigure}%
	\hspace{0.01\linewidth}%
	\begin{subfigure}{.495\linewidth}
		\centering
		\includegraphics[width=\linewidth]{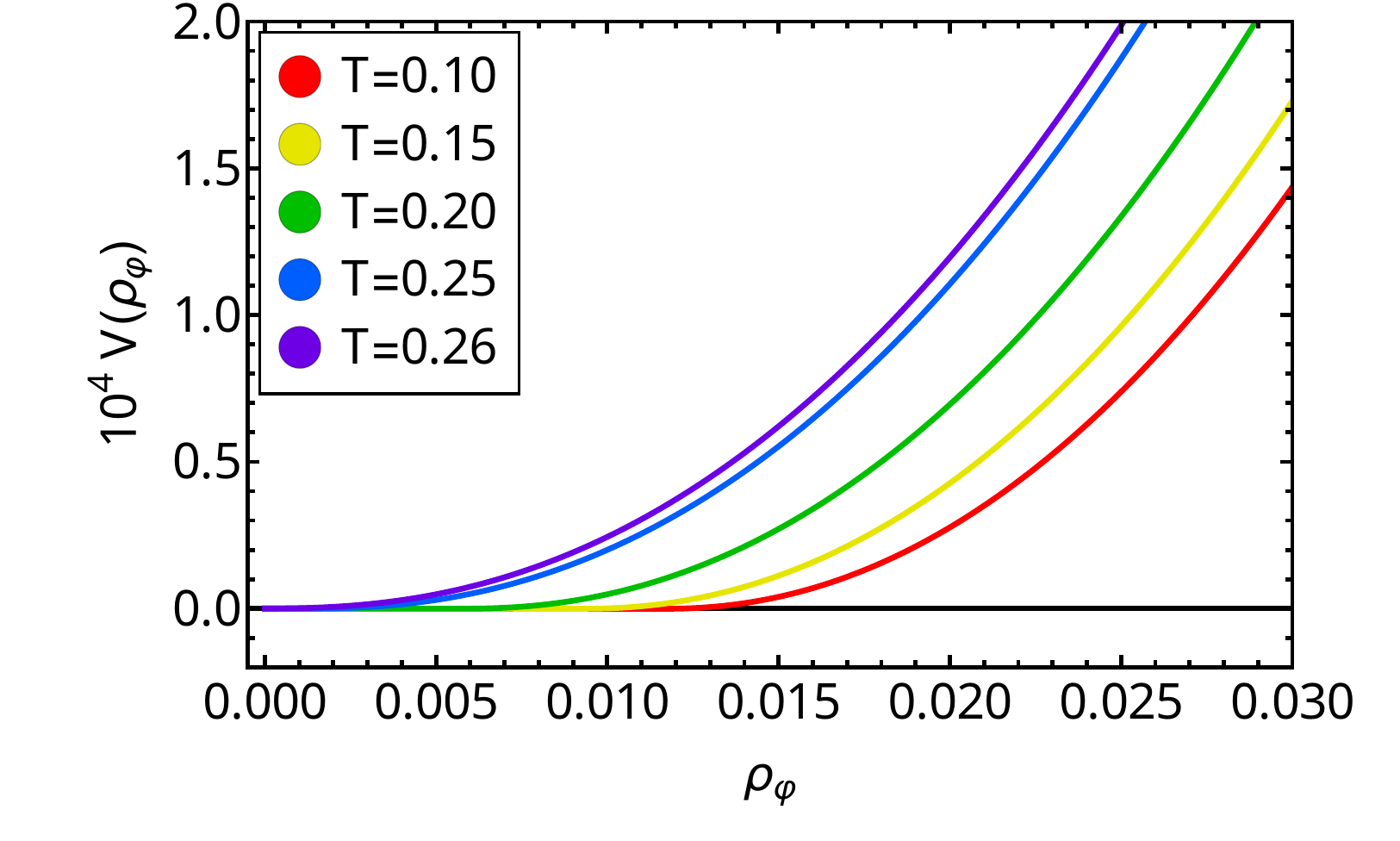}
		\caption{Potential for different temperatures at $k=0.002$. \hspace*{\fill}}
		\label{fig:potentialsd4T}
	\end{subfigure}
	\caption{Transformation of the field basis and potential in $d=4$ at different temperatures. We show results in terms of the coordinates at the initial cut-off scale $\Lambda$. The solution to the equation of motion \labelcref{eq:EoM} is indicated by the dashed lines. The critical temperature is given by $T_c \approx 0.26$. Hence the \textit{(purple)} line is located in the broken phase. All units are given in terms of the UV-cutoff $\Lambda = 1$.\hspace*{\fill}}
	\label{fig:physicsd4}
\end{figure*}
We have computed the effective potential around the phase transition with a temperature spacing of $\Delta T=10^{-3}$ for both the full $Z_\varphi(\rho_\varphi)$ dependence and in LPA$'$ down to a minimal RG-scale of $k_{\mathrm{min}} = 0.002$. From this investigation we can determine the location $T_c$ of the phase transition in the chiral limit in the present approach with flowing fields in comparison to the location $T_{c,\mathrm{LPA}'}$ in LPA$'$, 
\begin{align}
	T_{c} &=0.258(1)\,, \notag\\[1ex]
	T_{c,\mathrm{LPA}'} &= 0.259(1) \,.
\end{align}
The critical temperatures in both approximations agree within their error, which again provides a much desired systematic error estimate for a $T_c$-predictions in $T_{c,\mathrm{LPA}'}$ in the current model class.
The fine temperature resolution $\Delta T=10^{-3}$ or rather $\Delta T/T_c\approx 0.0039$ used in the present computation allows for a crude estimate of the size of the critical region. To begin with, we find that the used temperature spacing is not small enough for a reliable extraction of the critical exponents. Note that $\Delta T$ is readily reduced without problems given the numerical accuracy of the present framework. However, the present numerical approach is set-up for computing the effective potential in the phase structure, and is not tailor made for the extraction of critical exponents, which is best done within a fixed point analysis. When determining the scaling of $m_\pi^2 \propto |T-T_c|^\nu$, our estimate of the region that is \textit{dominated} by critical scaling, is 
\begin{align}
	 \left|1 -\frac{T_c}{T}\right| \lesssim 5 \times 10^{-3} \,.
\label{eq:CritScal}
\end{align}
Note that a fit with sub-leading scaling terms and regular terms may be compatible with our results for a far larger regime, but still the extraction of the critical exponents from $|1-T_c/T|\gg 5 \ 10^{-3}$ requires a rapidly increasing precision as the scaling contributions get suppressed. In any case we remark, that the critical exponent obtained from a naive interpolation of the potential in the regime \labelcref{eq:CritScal} is compatible with $\nu \approx 0.7 - 0.8$. Finally, we consider the anomalous dimension $\eta_{\phi}(\rho_0)$ on the equations of motion $\rho_0$, in the vicinity of the phase transition, see \Cref{fig:etaCrit}. Our computations support a critical value $\eta_{\phi,c}(\rho_0) \approx 0.042$.

\subsubsection{Some temperature dependent potentials}
We close the discussion of the thermal properties of the O(4) model within the flowing field approach with a discussion of the temperature dependence of the field transformation $\phi(\varphi)$ and the potential. The anomalous dimension in the symmetric phase at $T=0.270$ and in the broken phase at $T=0.100$ is depicted in \Cref{fig:dotPhi}. In the symmetric phase the map to the flowing fields encounters no significant non-linearities for $k\to 0$ as the latter originate in the non-convexity of the effective potential only present in the broken phase for $k\to 0$. \Cref{fig:flowingT} shows the field transformation at $k \to 0$ for different temperatures. The field-dependent wave function can be obtained from the coordinate reparametrisation at $k \to 0$ using
\begin{align}
	\phi = Z_\varphi (\rho_\varphi)^{1/2} \varphi \,,
\end{align}
and is depicted for various temperatures in \Cref{fig:fielddepZ}. The flat part of the coordinate transformation in \Cref{fig:flowingT} translates to the $\propto 1/\varphi$ behaviour that is indicated by the dashed lines in \Cref{fig:fielddepZ}. Most notably, $Z_\varphi$ is not maximal on the equations of motion $\phi_{\mathrm{EoM}}$ and thus $\partial_\rho Z_\varphi |_{\mathrm{EoM}} \neq 0$ in the broken phase, as suggested by the finite density approximation below \labelcref{eq:dynamHadron-eta}.
\Cref{fig:potentialsd4T} shows the potential at $k\to 0$ for various temperatures. 

\section{Summary and Outlook}
\label{sec:summary}

In this work we put forward an application of the generalised flow equation \labelcref{eq:GenFlow} to an O(4) theory, which implements the idea of an expansion of the effective action about the ground state of the theory. Technically, this is done by reducing the full dispersion with a field-dependent wave function $Z_\varphi[\varphi]$ of the fundamental scalar field $\varphi$ to a classical one at each flow step. This is achieved by a dynamical reparametrisation of the field $\varphi \to \phi_k(\varphi)$, introducing an emergent composite or flowing field via the differential reparametrisation $\dot \phi[\phi]$ with $Z_\varphi \to Z_\phi \equiv 1$ at each RG-step.

As a first application we have computed the effective potential in the three-dimensional theory in the broken phase within this optimised expansion scheme. The results have been compared to the commonly used LPA$'$ approximation scheme, in which the wave function $Z_\varphi$ is only computed on the cutoff-dependent solution of the equation of motion. Our results provide a non-trivial reliability check of the LPA$'$ approximation in the broken phase, which turns out to be quantitatively reliable in the present setup. However, the sizeable deviations of the solutions away from the equation of motion depicted in \Cref{fig:pot3} indicate, that higher correlation functions may differ in LPA$'$. For this reasons we also expect inaccuracies of the LPA$'$ scheme in the presence of highly dynamical flows. For instance, we hope to report in the near future on an application of flowing fields to shock-development processes, as e.g.~present at large densities and low temperatures in the QCD phase diagram.

Finally, we have applied the flowing field approach to the thermal phase transition of the four-dimensional $\phi^4$ theory. As for the three-dimensional case we found an impressive reliability of the LPA$'$ approximation for the observables considered. For example, the critical temperatures agreed in both approximations within their small respective error. This result has an immediate consequence for the systematic error estimate of state of the art fRG computations for the chiral phase transition up to $\mu_B/T \approx 4$: the current work supports the quantitative accuracy of the LPA$'$ scheme used e.g. in \cite{Fu:2019hdw} for the chiral part of QCD. 

The present approach can be readily implemented in the fRG approach to first principles QCD, leading to a flowing optimised expansion scheme around the cutoff-dependent ground state in the mesonic sector. This improves the truncation scheme at finite temperature and density in the quest for quantitative precision for high density QCD. The latter application also has to take into account potential spatial inhomogeneities such as a moat regime \cite{Pisarski:2021qof, Rennecke:2023xhc}. This requires momentum-dependent wave functions, and hence the implementation of a momentum-dependent flowing field transformation. We hope to report on respective results in the near future.


\begin{acknowledgments}

We thank Kevin Falls and Manfred Salmhofer for discussions. We thank Franz R. Sattler and Nicolas Wink for discussions and collaborations on related projects. This work is done within the fQCD-collaboration~\cite{fQCD}, and we thank the members for discussions. This work is funded by the Deutsche Forschungsgemeinschaft (DFG, German Research Foundation) under Germany’s Excellence Strategy EXC 2181/1 - 390900948 (the Heidelberg STRUCTURES Excellence Cluster) and the Collaborative Research Centre SFB 1225 - 273811115 (ISOQUANT). FI acknowledges support by the Studienstiftung des deutschen Volkes. 
\end{acknowledgments}


\appendix
\begingroup
\allowdisplaybreaks

\section{Diagrammatics of the modified flow equation}
\label{app:Zprojection}

In this section we outline the derivation of the modified flow of the wave function. The full flow of the two-point function reads with $R_{k,ij}=R\,\delta_{ij}$,
\begin{align}\nonumber 
	\partial_t \Gamma_{i j} =&\, \partial_t \Gamma_{i j}|_{\dot{\phi} =0}+ \Biggl[G_{nm,ij} \dot{\phi}_{n,m} \\ 
	&\hspace{-.5cm}+ G_{nm,i} \dot{\phi}_{m, nj} + G_{nm,j} \dot{\phi}_{m, ni} + G_{nm} \dot{\phi}_{m,rij}\Biggr] R
	 \notag \\[1ex]
	&\hspace{-.5cm}- 
	 \left(\dot{\phi}_n \Gamma_{n i j} + \dot{\phi}_{n,i} \Gamma_{n j} +\dot{\phi}_{n,j} \Gamma_{n i} + \dot{\phi}_{n,i j} \Gamma_{n}\right) \,, 
\label{eq:secondDeriv}
\end{align}
where we used the short-hand notation introduced in \labelcref{eq:GnNotation,eq:G2Notation} and below as well as 
\begin{align}
 \dot{\phi}^{(m)}_{i, \phi_{n_1}\cdots \phi_{n_m}} = \frac{ \delta \dot{\phi}_i }{\delta\phi_{n_1} \cdots \delta\phi_{n_m}}\,, 
\label{eq:dotphinNotation}
\end{align} 
as well as the short hand notation $ \dot{\phi}_{i, {n_1}\cdots n_m}$. In \labelcref{eq:dotphinNotation} we have dropped the momentum arguments. Note also that all quantities in \labelcref{eq:secondDeriv} depend on $k$. 

The first contribution is the standard 1PI Wetterich flow of the two-point function. The next line can also be thought of in terms of diagrams: the terms correspond to the 2-, 1- point-function and potential flows respectively, but the regulator term is replaced by $\partial_t R_k \to \dot{\phi}^{(m)} R_k$, where the superscript denoted the derivative and $m=1,2,3$. $\dot{\phi}^{(2)}$ has one outer leg, $\dot{\phi}^{(3)}$ two. 
We can use this picture to drop some of the terms. Since the reparametrisation $\dot{\phi}$ has no external momentum dependence, the diagram containing $\dot{\phi}^{(3)}$ drops after taking the momentum derivative.
 
Furthermore, all $\dot{\phi}_{,\phi \pi_i}$ contributions drop when evaluated at the expansion point $\phi_0=(\sqrt{\rho}, \bm 0)$. 
We can see this by a symmetry argument: The left hand side of \labelcref{eq:secondDeriv} is a function of the invariant $\rho$. Thus the right hand side is necessarily too, which implies that $\dot{\phi}^{(1)}$ is a function of $\rho$, whereas $\dot{\phi}$ and $\dot{\phi}^{(2)}$ are odd in $\phi$. Using the relation
\begin{align}\label{eq:rhoandphiderivs}
	\partial_{\phi_i} f(\rho) = \phi_i \ \partial_\rho f(\rho) \,,
\end{align}
for an arbitrary function $f(\rho)$, we find that 
\begin{align}
	\dot{\phi}_{\phi \pi_i}\vert_{\phi=\phi_0} = \pi_i \ \partial_\rho \frac{\delta \dot{\phi}}{\delta \phi} \vert_{\phi=\phi_0} = 0\,.
\end{align}
%

\section{Numerical setup}
\label{sec:NumericalSetup}

With the introduction of the RG-adapted, field-dependent flow of the anomalous dimension $\eta_{\phi,k}(\rho)$, \labelcref{eq:Exdotphi1}, we add an additional ordinary differential equation to the RG-time integration of the field-dependent effective potential $V_k(\rho)$. 
Technically, this ODE can be solved at every step of the RG-time integration and then reinserted into the flow. 
In doing this, we insert additional derivative terms to the evolution equation of the potential \labelcref{eq:flowV}, which affect the stability of the numerical scheme.

For simplicity we chose to insert the field dependent wave function iteratively. The first iteration step is initialised with $\eta_{\phi,0} = 0$:
\begin{itemize}
	\item \textit{(i)} The potential $V_{n}(\rho, k)$ is obtained by solving \labelcref{eq:flowV}, using $\eta_{\phi,n-1}(\rho,k)$. For details see \Cref{sec:finiteDiff}.
	\item \textit{(ii)} We use the potential $V_{n}(\rho, k)$ to obtain $\eta_{\phi,n}(\rho,k)$ from \labelcref{eq:Exdotphi1}. Details are given in \Cref{sec:statEq}.
\end{itemize}
\begin{figure}
	\centering
	\includegraphics[width=\linewidth]{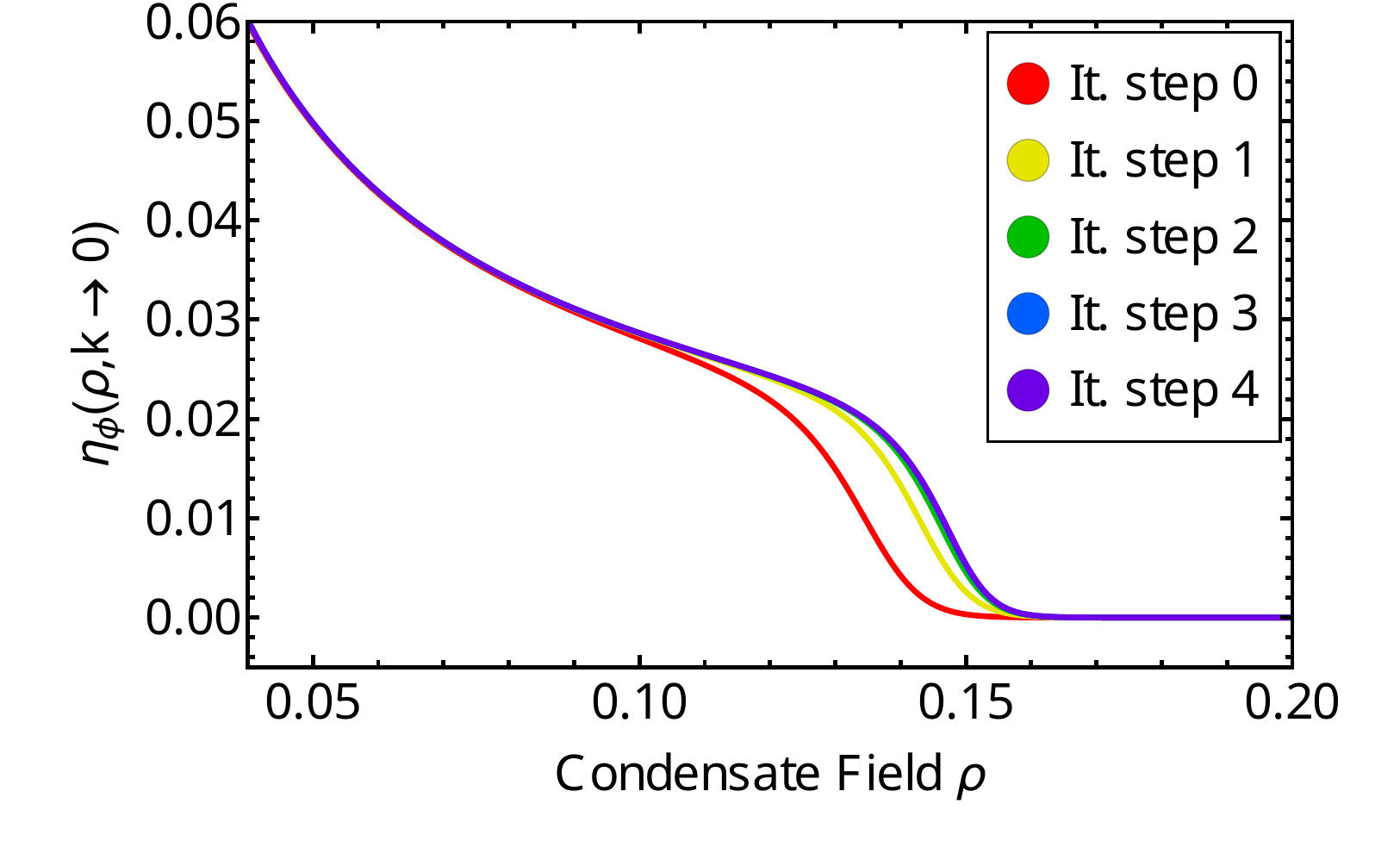}
	\caption{Iteration of the anomalous dimension $\eta_{\phi,n}$ in $d=3$ at $T=0.05$. The result is shown at $k \to 0$.\hspace*{\fill}}
	\label{fig:iteration}
\end{figure}
The iteration procedure converges around the equations of motion after 4 iteration steps, see e.g. \Cref{fig:iteration}.
The RG-time evolution is started at the UV-scale $k=\Lambda$ and we use small, discrete RG-steps $k \to k - \Delta k$ until the final RG-scale $k= k_{\mathrm{fin}} \approx 0$ is reached. The field dependence of both functions is discretised on an interval $\rho \in (0, \rho_{\mathrm{UV}}]$ using a simple equidistant grid, see \Cref{sec:finiteDiff}.

\subsection{Finite difference discretisation}\label{sec:finiteDiff}

The field dependent potential derivative $\partial_{{\rho}} V (\rho) =u(\rho)$ and anomalous dimension $\eta_{\phi}(\rho)$ are solved using a finite difference discretisation. This simple numerical scheme has previously been applied to the fRG in \cite{Ihssen:2023qaq}. The dependence on $\rho$ is discretised on a grid, using the notation
\begin{align}
	\label{eq:basic_grid}
	\{\rho_i\} = \left\{ \rho_1 = 0, \rho_2, \dots, \rho_{N_\rho} = \rho_\tinytext{max} \right\}
	\, ,
\end{align}
with $N = 512$.
We evaluate the functions on every grid-point using the notation $u_i = u(t, \rho_i)$ as well as $\eta_{\phi,i} =\eta_{\phi}(t, \rho_i)$. Derivatives are obtained from the difference to the next grid point. We distinguish between the upwind and downwind derivative operators,
\begin{align}
	\label{eq:deriv_operator}
	\mathcal{D}_{u(d)} f_i = \frac{f_{i\pm 1} - f_i}{\rho_{i\pm 1} - \rho_{i}}
	\, ,
\end{align}
where $f$ is the discretised function.

The flow of the potential \labelcref{eq:flowV} defines a second order differential equation, which is dependent on its first and second derivatives $\partial_{{\rho}} V(\rho)$ and $\partial^2_{{\rho}} V(\rho)$ respectively. It can be rewritten as a convection-diffusion equation along the lines of \cite{Grossi:2019urj, Grossi:2021ksl, Ihssen:2022xkr}, by taking a $\rho$ derivative. Hence we solve for $u=\partial_{{\rho}} V(\rho)$, with,
\begin{align}\label{eq:schemFlow}
	\partial_t u =  \partial_\rho \left(\rho \eta_{\phi} \, u + F \right)\,,
\end{align}
where the flow $F(u, u', \rho,\eta_{\phi},\eta_{\phi}')$ is given by the right hand side of \labelcref{eq:flowV}.
Within the finite difference scheme, one needs to consider the information flow to determine which derivative operators to use.
In previous works \cite{Ihssen:2023qaq, Grossi:2019urj} the RG-flows of the potential have been found to flow from high field values to smaller field values. Hence, the outer derivative  of the flow $F$ in \labelcref{eq:schemFlow} is discretised with a upwind derivative. To make for a central derivative of the diffusive part, we use a downwind derivative for $u'$. The anomalous dimension term is given by external input and strictly $<1$. Therefore, it does not change the direction of the flow and we can simply add it.

\subsection{Solving the RG-adapted flow}\label{sec:statEq}

We solve the RG-adapted flow \labelcref{eq:Exdotphi1} at each RG-scale. 
The flow is a first order ordinary differential equation (ODE), which reduces to solving a linear system of equations for $\eta_{\phi,i}$
\begin{align}
	\eta_{\phi,i} = \,{\cal T}_{\eta_\phi}(u_i,\mathcal{D}_{u} u_i ,\rho_i;T) \left( 1 - \frac{	\eta_{\phi,i} + 2 \rho_i \mathcal{D}_{u} \eta_{\phi,i} }{d+1}\right) \,,
\end{align}
where we have the threshold function
\begin{align}
	{\cal T}_{\eta_\phi}(\bar V',\bar V'',\rho) = 4 A_d\, \rho (\bar V'')^2 \mathcal{BB}_{(2,2)}\,.
	\label{eq:Threshold}
\end{align}
We are exclusively using the upwind derivative operator in solving this equation for $\eta_\phi$. This is due to the boundary condition $\eta_{\phi,1} = 0$, which originates from ${\cal T}_{\eta_\phi}(\bar V',\bar V'',0) = 0$. 

This simple set-up is preferable to solving an ODE for $\eta_{\phi}$ by integrating the derivative $\partial_{{\rho}}\rho \eta_{\phi}$, since it only appears suppressed by the threshold function.

The iteration procedure of the anomalous dimension does not converge at very small field values in the broken phase, which is due to the finite difference discretisation and a very steep slope around $\rho = 0$. This is corrected numerically, by freezing in the value of  $\eta_\phi$ for the first 10 grid-point at some finite $k>0$ as soon as the oscillations begin. Due to the direction of the information flow, this procedure does not back-propagate to the equations of motion. \Cref{fig:iteration} shows the convergence of the iteration procedure, the frozen points are located at smaller values of the field $\rho$.

\section{Choice of regulator}
\label{app:regulator}

For the derivation of flows in the O(N)-theory we use a spatial three-dimensional flat regulator, \cite{Litim:2000ci, Litim:2001up, Litim:2006ag}
\begin{subequations}
\label{eq:FlatReg}
\begin{align}
	R_{k,ij} (\bm{q}) &= R\, \delta_{ij} \,,\quad R= Z_\varphi(\varphi_0) \bm{q}^2 \, r \left(\bm{q}^2/k^2\right)\,, 
\label{eq:Regulators}
\end{align}
with the shape function 
\begin{align}
	r (x) &= \left(\frac{1}{x} -1 \right) \Theta(1-x) \,. 
\label{eq:RegShapefunction}
\end{align}
\end{subequations}
This regulator is a common choice for thermal fRG flows in LPA and LPA$'$, owing to the convenient analytical form of the respective flow equations for the effective potential and the wave function. Moreover, at $T\to \infty$ it is the optimal regulator \cite{Litim:2000ci, Litim:2001up, Pawlowski:2005xe}, while at $T=0$ its four-dimensional version is optimal. Functional optimisation, \cite{Pawlowski:2005xe}, beyond LPA leads to a smoothened version of \labelcref{eq:Regulators}, however, a combination of the flowing field approach with functional optimisation goes beyond the scope of the present work. 

Here, we choose $\varphi_0$ as the solution to the equation of motion \labelcref{eq:EoM}. With the flowing fields introduced in \Cref{sec:RG-kernel}, the wave function factor is simply $Z_\phi(\phi_0) = 1$ for all $k$ and drops out in \labelcref{eq:Regulators} as it does in the effective action. Seemingly, this LPA-type form of the effective action suggests \labelcref{eq:FlatReg} as an optimal regulator (at $T=0$). However, inserted into the functional optimisation condition from \cite{Pawlowski:2005xe} the $\dot\phi$-terms in \labelcref{eq:GenFlow} enforce a smoothening tantamount to that in the explicit presence of the wave function.

\section{Threshold functions}
\label{app:thrs}

The threshold function $\mathcal{BB}_{22}$ is the typical expression for a single loop expression containing two propagators of different bosonic species. It is computed in terms of the scalar part of the propagator with a flat cutoff, see \Cref{app:regulator} and is given by, 
\begin{align}\label{eq:ScalPProp}
	G_b (q, \bar m^2) &= \frac{1}{\left(q_0/k\right)^2 + 1 + \bar m^2} \,.
\end{align}
Here we introduce the dimensionless masses $\bar m = m/k$ for convenience.
The Matsubara frequency reads $q_0 = 2 j\pi T $, with $j \in \mathbb{Z}$. The threshold function is then defined as,
\begin{align}
	\mathcal{BB}_{22}(&\bar m_{1}^2,\bar m_{2}^2; T)\notag \\[1ex] &= \frac{T}{k} \sum_{j} \left( G_b (q, \bar m_{1}^2) \right)^{2} \left( G_b (q, \bar m_{2}^2) \right)^{2} \,.
\label{eq:BB22}
\end{align}
where $\bar m_1$ and $\bar m_2$ are the dimensionless masses of the two different field species.
Furthermore, we can infer the threshold function for two propagators of each species, by taking derivatives with respect to the dimensionless masses,
\begin{align}\label{eq:LoopDerivs}
\mathcal{BB}_{22}(&\bar m_{1}^2,\bar m_{2}^2; T)= \partial_{\bar m_{1}^2} \partial_{\bar m_{2}^2} \mathcal{BB}_{11}(\bar m_{1}^2,\bar m_{2}^2; T) \,,
\end{align}
Finally we perform the Matsubara summation for $\mathcal{BB}_{11}$, which is given by,
\begin{align}
	\mathcal{BB}_{(1,1)}(\bar m_{1}^2 &,\bar m_{2}^2; T) = -\frac{1}{2 ( \bar m^2_{1}- \bar m^2_{2})}\notag \\[1ex]
	&\left[\frac{\coth\left(\frac{k E_{1}}{2T}\right)}{E_{1}}
	-	\frac{\coth\left(\frac{k E_{2}}{2T}\right)}{E_{2}}\right]\,,
\end{align}
with the dispersion relation $E_i = \sqrt{1 + \bar m_i^2}$.

\endgroup

\bibliographystyle{apsrev4-2}
\bibliography{references}
\end{document}